\documentclass[preprint,showpacs,amsmath,amssymb]{revtex4}

\usepackage{graphicx}
\usepackage{dcolumn}
\usepackage{bm}

\begin{document}

\title{Phenomenological analysis of the double pion production in nucleon-nucleon collisions up to 2.2 GeV}

\author{Xu Cao$^{1,3,5}${\footnote{Electronic address: caoxu@impcas.ac.cn}} }
\author{Bing-Song Zou$^{2,3,4}$}
\author{Hu-Shan Xu$^{1,3,4}$}

\affiliation{$^1$Institute of Modern Physics, Chinese Academy of
Sciences, Lanzhou 730000, China\\
$^2$Institute of High Energy Physics, Chinese Academy of Sciences,
Beijing 100049, China\\
$^3$Theoretical Physics Center for Sciences Facilities, Chinese
Academy of Sciences,
Beijing 100049, China\\
$^4$Center of Theoretical Nuclear Physics, National
Laboratory of Heavy Ion Collisions, Lanzhou 730000, China\\
$^5$Graduate University of Chinese Academy of Sciences, Beijing
100049, China}

\begin{abstract}
With an effective Lagrangian approach, we analyze several $NN \to
NN\pi\pi$ channels by including various resonances with mass up to
1.72 GeV. For the channels with the pion pair of isospin zero, we
confirm the dominance of $N^*(1440)\to N\sigma$ in the near
threshold region. At higher energies and for channels with the final
pion pair of isospin one, we find large contributions from
$N^*(1440)\to \Delta\pi$, double-$\Delta$, $\Delta(1600) \to
N^*(1440)\pi$, $\Delta(1600) \to \Delta\pi$ and $\Delta(1620) \to
\Delta\pi$. There are also sizeable contributions from $\Delta \to
\Delta\pi$, $\Delta \to N\pi$, $N \to \Delta\pi$ and nucleon pole at
energies close to the threshold. We well reproduce the total cross
sections up to beam energies of 2.2 GeV except for the $pp\to
pp\pi^0\pi^0$ channel at energies around 1.1 GeV and our results
agree with the existing data of differential cross sections of $pp
\to pp\pi^+\pi^-$, $pp \to nn\pi^+\pi^+$ and $pp \to pp\pi^0\pi^0$
which are measured at CELSIUS and COSY.
\end{abstract}
\maketitle{}

\section{INTRODUCTION}

Double pion production in both pion- and photo-induced reactions has
been an intriguing field to study baryon spectrum and given insight
to the properties of strong interaction~\cite{reviewlee,schneider}.
These reactions close to threshold are also an interesting area to
test chiral symmetry and have been extensively explored
experimentally~\cite{twopiphoto} and theoretically~\cite{twopicpt}.
Recently the double pion production in the electro-production off
protons has advanced a important step~\cite{twopielectro}. All
essential contributions are identified from the data and the major
isobar channels are well determined. On the other hand, as another
fascinating platform for studying resonances properties, double pion
production in nucleon-nucleon collisions has been accurately
measured at the facilities of CELSIUS and COSY in the past few
years, and the comprehensive data of various differential cross
sections are obtained up to beam energies 1.3
GeV~\cite{celsius,dataIJMPA,skorodko,cosy,data2000,otherdata}.
However, on the theoretical side, the study of this reaction is
scarce.  The state-of-art one is still the Valencia model
calculation~\cite{Alvarez} of more than ten years ago after some
much earlier calculations of one pion exchange (OPE)
model~\cite{Ferrari} of more than 45 years ago. Thus a more
comprehensive analysis matching the modern data is very necessary.

The early OPE model, which mainly focused on the old data at beam
energies of 2.0 GeV and 2.85 GeV~\cite{olddata}, included two types
of diagrams with the final two pions produced from a single and two
baryon line(s), respectively. It used the amplitudes of $\pi N
\to\pi\pi N$ and $\pi N \to\pi N$ extracted from limited data and
the off-shell corrections were considered under several assumptions.
It did not account for the explicit production mechanisms of double
pion and other exchanged meson besides $\pi$-meson. The Valencia
model is characteristic by the the dominance of $N^*(1440)\to
N\sigma$ in the near threshold region in the isospin allowed
channels while the double-$\Delta$ and $N^*(1440)\to \Delta\pi$ rise
up at higher energies and in channels where $N^*(1440)\to N\sigma$
is forbidden by the isospin conservation. Recently, the
experimental data~\cite{celsius,cosy} confirm the predicted behavior
close to threshold and this makes $pp \to pp\pi^+\pi^-$ and $pp \to
pp\pi^0\pi^0$ good places to study $N^*(1440)$ whose structure is
still controversial. Current data seem to show a weaker
$N^*(1440)\to \Delta\pi$ than that listed in Particle Data
Group~\cite{pdg2008} and support the explanation of $N^*(1440)$ as
the monopole excitation of the nucleon. Contrarily, the case is much
more complicated at higher energies. The dominance of the
double-$\Delta$ mechanism in the Valencia model results in that the
total cross section of $pp \to pp\pi^0\pi^0$ is about a factor of four
larger than that of $pp \to nn\pi^+\pi^+$, while the new exclusive
and the old bubble-chamber data are consistent to conclude an
approximate equal value of these two channels. The isospin
decomposition unambiguously reveal that more isospin 3/2 resonances
besides $\Delta$ is required to explain the
data~\cite{celsius,skorodko}, and this is also the reason that the
Valencia model including simply the $N^*(1440)$ and $\Delta$
achieved merely a rough agreement in most channels. Indeed, at
higher energies the contribution from higher lying resonances,
especially those having large double pion decay channels, should
become relevant. The recent detailed measurements performed by
CELSIUS and COSY make the further exploration of these problems
possible.

In the present work, we try to incorporate the resonances with mass
up to 1.72 GeV in an effective Lagrangian model with the motivation
to give a reasonable explanation to the six isospin channels of $NN
\to NN\pi\pi$ simultaneously and get better understanding of
dynamics for this kind of reactions. Our paper is organized as
follows. In Sect.~\ref{formalism}, we present the formalism and
ingredients in our computation. The numerical results and discussion
are demonstrated in Sect.~\ref{discussion} and a brief summary is
given in Sect.~\ref{summary}.

\section{FORMALISM AND INGREDIENTS} \label{formalism}

\begin{figure}[htbp]
  \begin{center}
 {\includegraphics*[scale=0.8]{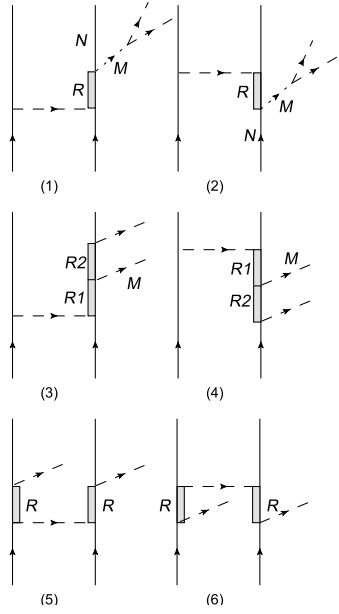}}
\caption{Feynman diagrams for $NN \rightarrow NN\pi\pi$. The solid,
dashed and dotted lines stand for the nucleon, mesons and
intermediate $\sigma$(or $\rho$)-meson. The shading histograms
represent the intermediate resonances or nucleon poles. In the text,
we use $R\to NM$, $R1\to R2M$ and double-$R$ to label (1)(2), (3)(4)
and (5)(6), respectively.} \label{fdg}
  \end{center}
\end{figure}

We consider nearly all possible Feynman diagrams as depicted in Fig.
1, and exchanged diagrams are also included. We use the commonly
used interaction Lagrangians for $\pi NN$, $\pi \Delta\Delta$, $\eta
NN$, $\sigma NN$ and $\rho NN$ couplings,
\begin{equation}
{\cal L}_{\pi N N} = - \frac{f_{\pi N N}}{m_{\pi}} \overline{N}
\gamma_5 \gamma_{\mu} \vec\tau \cdot \partial^{\mu} \vec{\pi} N ,
\label{pin}
\end{equation}
\begin{equation}
{\cal L}_{\pi \Delta \Delta} = \frac{f_{\pi \Delta \Delta}}{m_{\pi}}
\overline{\Delta}^{\nu} \gamma_5 \gamma_{\mu} \vec\tau \cdot
\partial^{\mu} \vec{\pi} \Delta_{\nu} + h.c.,
\end{equation}
\begin{equation}
{\cal L}_{\eta N N} = -i g_{\eta N N} \overline{N} \gamma_5 \eta N,
\label{etan}
\end{equation}
\begin{equation}
{\cal L}_{\sigma N N} = g_{\sigma N N}  \overline{N} \sigma N,
\label{sign}
\end{equation}
\begin{equation}
{\cal L}_{\rho N N} = -g_{\rho N N}
\overline{N}(\gamma_{\mu}+\frac{\kappa}{2m_N} \sigma_{\mu \nu}
\partial^{\nu})\vec\tau \cdot \vec\rho^{\mu} N. \label{rhon}
\end{equation}
At each vertex a relevant off-shell form factor is used. In our
caculation, we take the same form factors as that used in the Bonn
potential model~\cite{bonn},
\begin{equation}
F^{NN}_M(k^2_M)=\left(\frac{\Lambda^2_M-m_M^2}{\Lambda^2_M-
k_M^2}\right)^n,
\end{equation}
with n=1 for $\pi$- and $\eta$-meson and n=2 for $\rho$-meson.
$k_M$, $m_M$ and $\Lambda_M$ are the 4-momentum, mass and cut-off
parameters for the exchanged meson, respectively. The coupling
constants  and the cutoff parameters are taken as
~\cite{bonn,tsushima,singlepi}: $f_{\pi NN}^2/4\pi = 0.078$, $g_{\eta
NN}^2/4\pi = 0.4$, $g_{\sigma NN}^2/4\pi = 5.69$,
$g_{\rho NN}^2/4\pi = 0.9$, $\Lambda_{\pi}$ = $\Lambda_{\eta}$ = 1.0 GeV,
$\Lambda_{\sigma}$ = 1.3 GeV, $\Lambda_{\rho}$ =
1.6 GeV, and $\kappa$ = 6.1. We use $f_{\pi \Delta \Delta}=4f_{\pi
NN}/5$ from the quark model~\cite{reviewlee,Alvarez}. The mass and
width of $\sigma$-meson are adopted as 550 MeV and 500 MeV, respectively.

We include all $N^*$ and $\Delta^*$ resonances with spin-parity
$1/2^\pm$, $3/2^\pm$, $5/2^\pm$ and mass up to 1.72 GeV listed in
Particle Data Group (PDG) tables~\cite{pdg2008}. The resonances with
further higher masses are expected to give negligible contributions
in the energy region considered here and their two pion branching
ratios have large uncertainties, so we do not include them at
present. The effective Lagrangians for the relevant resonance
couplings are~\cite{zoucoupling,ouyang},
\begin{equation}
{\cal L}^{1/2^+}_{\pi N R} = g_{\pi NR} \overline{N} \gamma_5
\gamma_{\mu} \vec\tau \cdot \partial^{\mu} \vec{\pi} R + h.c.,
\end{equation}
\begin{equation}
 {\cal L}^{1/2^+}_{\eta N R} = g_{\eta NR} \overline{N} \gamma_5 \eta R + h.c.,
\end{equation}
\begin{equation}
{\cal L}^{1/2^+}_{\sigma N R} = g_{\sigma N R}  \overline{N} \sigma
R + h.c.,
\end{equation}
\begin{equation}
{\cal L}^{1/2^+}_{\rho N R} = g_{\rho N R}  \overline{N}
\gamma_{\mu} \vec\tau \cdot \vec{\rho^{\mu}} R + h.c.,
\end{equation}
\begin{equation}
{\cal L}^{1/2^+}_{\pi \Delta R} = g_{\pi \Delta R}
\overline{\Delta}_{\mu} \vec\tau \cdot \partial^{\mu} \vec{\pi} R +
h.c.,
\end{equation}
\begin{equation}
{\cal L}^{1/2^-}_{\pi N R} = g_{\pi NR} \overline{N} \vec\tau \cdot
\vec{\pi} R + h.c.,
\end{equation}
\begin{equation}
{\cal L}^{1/2^-}_{\eta N R} = g_{\eta NR} \overline{N} \eta R +
h.c.,
\end{equation}
\begin{equation}
{\cal L}^{1/2^-}_{\rho N R} = g_{\rho NR}\overline{N} \gamma_5
(\gamma_{\mu}-\frac{q_{\mu} \not \! q}{q^2}) \vec\tau \cdot
\vec{\rho^{\mu}} R + h.c.,
\end{equation}
\begin{equation}
{\cal L}^{1/2^-}_{\pi \Delta R} = g_{\pi \Delta R}
\overline{\Delta}_{\mu} \gamma_5 \vec\tau \cdot \partial^{\mu}
\vec{\pi} R + h.c.,
\end{equation}
\begin{equation}
{\cal L}^{3/2^+}_{\pi N R} = g_{\pi NR} \overline{N} \vec\tau \cdot
\partial^{\mu} \vec{\pi} R_{\mu} + h.c.,
\end{equation}
\begin{equation}
{\cal L}^{3/2^+}_{\eta N R} = g_{\eta NR} \overline{N}
\partial^{\mu} \eta R_{\mu} + h.c.,
\end{equation}
\begin{equation}
{\cal L}^{3/2^+}_{\rho N R} = g_{\rho NR} \overline{N} \gamma_5
\vec\tau \cdot \vec{\rho^{\mu}} R_{\mu} + h.c.,
\end{equation}
\begin{equation}
{\cal L}^{3/2^+}_{\pi \Delta R} = g_{\pi \Delta R}
\overline{\Delta}^{\mu} \gamma_5 \vec\tau \cdot \vec{\pi} R_{\mu} +
h.c.,
\end{equation}
\begin{equation}
{\cal L}^{3/2^+}_{\pi N^*(1440) R} = g_{\pi N^*R} \overline{N^*}
\vec\tau \cdot
\partial^{\mu} \vec{\pi} R_{\mu} + h.c.,
\end{equation}
\begin{equation}
{\cal L}^{3/2^-}_{\pi N R} = g_{\pi NR} \overline{N} \gamma_5
\gamma_{\mu} \vec\tau \cdot \partial^{\mu} \partial^{\nu} \vec{\pi}
R_{\nu} + h.c.,
\end{equation}
\begin{equation}
{\cal L}^{3/2^-}_{\rho N R} = g_{\rho NR} \overline{N} \vec\tau
\cdot \vec{\rho^{\mu}} R_{\mu} + h.c.,
\end{equation}
\begin{equation}
{\cal L}^{3/2^-}_{\pi \Delta R} = g_{\pi \Delta R}
\overline{\Delta}^{\mu} \gamma_{\nu} \vec\tau \cdot \partial^{\nu}
\vec{\pi} R_{\mu} + h.c.,
\end{equation}
\begin{equation}
{\cal L}^{5/2^+}_{\pi N R} = g_{\pi NR} \overline{N} \gamma_5
\gamma_{\mu} \vec\tau \cdot \partial^{\mu} \partial^{\nu}
\partial^{\lambda} \vec{\pi} R_{\nu\lambda} + h.c.,
\end{equation}
\begin{equation}
{\cal L}^{5/2^+}_{\rho N R} = g_{\rho NR} \overline{N}
\left(p_N^{\mu}-\frac{p_N\cdot p_R p_R^{\mu}  }{p^2_R}\right)
\vec\tau \cdot \vec{\rho^{\nu}} R_{\mu\nu} + h.c.,
\end{equation}
\begin{equation}
{\cal L}^{5/2^+}_{\sigma N R} = g_{\sigma NR} \overline{N}
\partial^{\mu} \partial^{\nu} \sigma R_{\mu\nu} + h.c.,
\end{equation}
\begin{equation}
{\cal L}^{5/2^+}_{\pi \Delta R} = g_{\pi \Delta R}
\overline{\Delta}_{\mu} \vec\tau \cdot
\partial^{\mu} \partial^{\nu}
\partial^{\lambda} \vec{\pi} R_{\nu\lambda} + h.c.,
\end{equation}
\begin{equation}
{\cal L}^{5/2^-}_{\pi N R} = g_{\pi NR} \overline{N} \vec\tau \cdot
\partial^{\mu} \partial^{\nu} \vec{\pi} R_{\mu\nu} + h.c.,
\end{equation}
\begin{equation}
{\cal L}^{5/2^-}_{\pi \Delta R} = g_{\pi \Delta R}
\overline{\Delta}_{\mu} \gamma_5 \vec\tau \cdot
\partial^{\mu} \partial^{\nu}
\partial^{\lambda} \vec{\pi} R_{\nu\lambda} + h.c.,
\end{equation}
For the Resonance-Nucleon-Meson vertices, form factors with the
following form are used:
\begin{equation}
F^{RN}_M(k^2_M)=\left(\frac{\Lambda^{*2}_M-m_M^2}{\Lambda^{*2}_M- k_M^2}\right)^n,
\end{equation}
with n=1 for $N^*$ resonances and n=2 for $\Delta$ resonances. We
employ $\Lambda^*_{\pi}$ = $\Lambda^*_{\sigma}$ = $\Lambda^*_{\eta}$
= $\Lambda^*_{\rho}$ = 1.0 for all resonances except
$\Lambda^*_{\pi}$ = 0.8 for $\Delta^*(1600)$. We also use
Blatt-Weisskopf barrier factors $B(Q_{N^*\Delta\pi})$ in the
$N^*(1440)$-$\Delta$-$\pi$ vertices~\cite{zouform},
\begin{equation}
B(Q_{N^*\Delta\pi})=\sqrt{\frac{P^2_{N^*\Delta\pi}+Q^2_{0}}{Q^2_{N^*\Delta\pi}+Q^2_{0}}},
\end{equation}
Here $Q_0$ is the hadron ¡°scale¡± parameter $Q_0=0.197327/R$ GeV/c,
where R is the radius of the centrifugal barrier in the unit of fm
and is tuned to be 1.5 fm to fit the data. $Q_{N^*\Delta\pi}$ and
$P_{N^*\Delta\pi}$ is defined as,
\begin{equation}
Q^2_{N^*\Delta\pi}=\frac{(s_N^*+s_\Delta-s_\pi)^2}{4s_N^*}-s_\Delta,
\end{equation}
\begin{equation}
P^2_{N^*\Delta\pi}=\frac{(m^{2}_{N^*}+m^2_\Delta-m^2_\pi)^2}{4m^{2}_{N^*}}-m^2_\Delta,
\end{equation}
with $s_x$ being the invariant energy squared of $x$ particle.
Because the mass of $\sigma$-meson is near the two-$\pi$ threshold,
the following Lagrangians and form factor are employed for the
$\sigma$-$\pi$-$\pi$ vertex~\cite{reviewlee,schneider,juelich},
\begin{equation}
{\cal L}_{\sigma\pi\pi} =
g_{\sigma\pi\pi}\partial^{\mu}\vec{\pi}\cdot
\partial_{\mu} \vec{\pi} \sigma,
\end{equation}
\begin{equation}
{\cal L}_{\rho\pi\pi} = g_{\rho\pi\pi}\vec{\pi}\times
\partial_{\mu} \vec{\pi}\cdot \vec{\rho^{\mu}},
\end{equation}
\begin{equation}
F^{\pi\pi}_{\sigma}(\vec{q}^{2})=\left(\frac{\Lambda^{2}+\Lambda_0^{2}}{\Lambda^{2}+
\vec{q}^{2}}\right)^2,
\end{equation}
where $\vec{q}$ is the relative momentum of the emitted
$\pi$-mesons. We use $\Lambda = 0.8$ GeV and $\Lambda_0^{2} = 0.12$
GeV$^2$ to normalize this form factor to unity when $\pi$- and
$\sigma$- meson are all on-shell. The decay width of $\sigma\to\pi\pi$
and $\rho\to\pi\pi$ yield $g^2_{\sigma\pi\pi}$ = $6.06$ and $g^2_{\rho\pi\pi}$ = $2.91$.


\begin{table}[htbp]
\caption{Relevant parameters used in our calculation. The masses,
widths and branching ratios (BR) are taken from central values of
PDG~\cite{pdg2008} except the BR for $N^*(1440)\to\Delta\pi$.}
\label{coupling}
\begin{center}
\begin{tabular}{cccccc}
\hline\hline
Resonance   &Pole Position &BW Width & Decay Mode & Decay Ratio &$g^2/4\pi$\\
\hline
$\Delta^*(1232) P_{33}$ & (1210, 100)& 118& $N\pi$& 1.0& 19.54\\
$N^*(1440) P_{11}$  & (1365, 190)& 300& $N\pi$& 0.65& 0.51\\
     &  &  & $N\sigma$& 0.075&3.20\\
     &  &  & $\Delta\pi$& 0.135&4.30\\
$N^*(1520) D_{13}$  & (1510, 110)& 115& $N\pi$& 0.6& 1.73\\
     &  &  & $N\rho$& 0.09&1.32\\
     &  &  & $\Delta\pi$& 0.2&0.01\\
$N^*(1535)S_{11}$  & (1510, 170)& 150& $N\pi$& 0.45& 0.037\\
     &  &  & $N\eta$& 0.525&0.34\\
     &  &  & $N\rho$& 0.02& 0.15\\
$\Delta^*(1600) P_{33}$  & (1600, 300)& 350& $N\pi$& 0.175& 1.09\\
     &  &  & $\Delta\pi$& 0.55&59.9\\
     &  &  & $N^*(1440)\pi$& 0.225&289.1\\
$\Delta^*(1620) S_{31}$  & (1600, 118)& 145& $N\pi$& 0.25& 0.06\\
     &  &  & $N\rho$& 0.14&0.37\\
     &  &  & $\Delta\pi$& 0.45&83.7\\
$N^*(1650)S_{11}$  & (1655, 165)& 165 & $N\pi$ &0.775 & 0.06\\
     &  &  & $N\eta$& 0.065&0.026\\
     &  &  & $N\rho$& 0.08&0.011\\
     &  &  & $\Delta\pi$& 0.04&0.063\\
$N^*(1675) D_{15}$  & (1660, 135)& 150 & $N\pi$ &0.4 & 2.16\\
     &  &  & $\Delta\pi$& 0.55&3077.5\\
$N^*(1680) F_{15}$  & (1675, 120)& 130 & $N\pi$ &0.675 & 5.53\\
     &  &  & $N\sigma$& 0.125&4.45\\
     &  &  & $N\rho$& 0.09&0.32\\
     &  &  & $\Delta\pi$& 0.1&9.39\\
\hline\hline
\end{tabular}
\end{center}
\end{table}
\begin{table}[htbp]
\caption{Table~\ref{coupling} continued.} \label{table}
\begin{center}
\begin{tabular}{ccccccc}
\hline\hline
Resonance   &Pole Position &BW Width & Decay Mode & Decay Ratio &$g^2/4\pi$ \\
\hline
$N^*(1700) D_{13}$  & (1680, 100)& 100& $N\pi$& 0.1& 0.075\\
     &  &  & $N\rho$& 0.07& 0.043\\
     &  &  & $\Delta\pi$& 0.04&0.003\\
$\Delta^*(1700) D_{33}$  & (1650, 200)& 300& $N\pi$& 0.15& 1.02\\
     &  &  & $N\rho$& 0.125& 0.69\\
     &  &  & $\Delta\pi$& 0.45&0.072\\
$N^*(1710) P_{11}$  & (1720, 230)& 100 & $N\pi$ &0.15 & 0.012\\
     &  &  & $N\eta$& 0.062&0.042\\
     &  &  & $N\sigma$& 0.25&0.085\\
     &  &  & $N\rho$& 0.15&36.1\\
     &  &  & $\Delta\pi$& 0.275&0.12\\
$N^*(1720) P_{13}$  & (1675, 195)& 200 & $N\pi$ &0.15 & 0.12\\
     &  &  & $N\eta$& 0.04&0.28\\
     &  &  & $N\rho$& 0.775&190.7\\
\hline\hline
\end{tabular}
\end{center}
\end{table}

The form factor for the resonance, $F_{R}(q^2)$, is taken as,
\begin{equation}
F_{R}(q^2)=\frac{\Lambda_R^{4}}{\Lambda_R^{4} + (q^2-M^2_R)^2},
\end{equation}
with $\Lambda_R$ = 1.0 GeV. The same type of form factors are also
applied to the nucleon pole with $\Lambda_N$ = 0.8 GeV. The
propagators of the exchanged meson, nucleon pole and resonance can
be written as~\cite{bonn,tsushima},
\begin{equation}
G_{\pi/\eta}(k_{\pi/\eta})=\frac{i}{k^{2}_{\pi/\eta}-m^{2}_{\pi/\eta}},
\end{equation}
\begin{equation}
G_{\sigma}(k_{\sigma})=\frac{i}{k^{2}_{\sigma}-m^{2}_{\sigma}+im_{\sigma}\Gamma_{\sigma}},
\end{equation}
\begin{equation}
G^{\mu\nu}_{\rho}(k_{\rho})=-i\frac{g^{\mu\nu}-k_{\rho}^{\mu}
k_{\rho}^{\nu}/k_{\rho}^{2}}{k^{2}_{\rho}-m^{2}_{\rho}},
\end{equation}
\begin{equation}
G_{N}(q)=\frac{ -i(\not \! q +m_{N})}{q^2-m^2_{N}}.
\end{equation}
\begin{equation}
G^{1/2}_{R}(q)=\frac{ -i(\not \! q
+M_{R})}{q^2-M^2_{R}+iM_{R}\Gamma_{R}}.
\end{equation}
\begin{equation}
G^{3/2}_{R}(q)=\frac{ -i (\not\! q + M_R)
G_{\mu\nu}(q)}{q^2-M^2_{R}+iM_{R}\Gamma_{R}}.
\end{equation}
\begin{equation}
G^{5/2}_{R}(q)=\frac{ -i (\not\! q +
M_R)G_{\mu\nu\alpha\beta}(q)}{q^2-M^2_{R}+iM_{R}\Gamma_{R}}.
\end{equation}
Here $\Gamma_{R}$ is the total width of the corresponding resonance,
and $G_{\mu\nu}(q)$ and $G_{\mu \nu \alpha \beta}(q)$ is defined as,
\begin{equation}
G_{\mu \nu}(q) = - g_{\mu \nu} + \frac{1}{3} \gamma_\mu \gamma_\nu +
\frac{1}{3 M_R}( \gamma_\mu q_\nu - \gamma_\nu q_\mu) + \frac{2}{3
M^2_R} q_\mu q_\nu,
\end{equation}
\begin{eqnarray}
G_{\mu \nu \alpha \beta}(q) = &-& \frac{1}{2}(\tilde
g_{\mu\alpha}\tilde g_{\nu\beta}+\tilde g_{\mu\beta}\tilde
g_{\nu\alpha} ) + \frac{1}{5}\tilde
g_{\mu\nu}\tilde g_{\alpha\beta}\\
&-&\frac{1}{10}(\tilde\gamma_\mu\tilde\gamma_\alpha\tilde
g_{\nu\beta}+\tilde\gamma_\nu\tilde\gamma_\beta\tilde
g_{\mu\alpha}+\tilde\gamma_\mu\tilde\gamma_\beta\tilde
g_{\nu\alpha}+\tilde\gamma_\nu\tilde\gamma_\alpha\tilde g_{\mu\beta}
) , \label{pmunualphabeta}
\end{eqnarray}
\begin{equation}
\tilde g_{\mu\nu}(q) = -g_{\mu\nu}+{q_\mu q_\nu\over M^2_R}, \quad
\tilde\gamma_\mu = -\gamma_\mu + {\not\! q q_\mu\over M^2_R}.
\end{equation}
Because constant width is used in the Breit-Wigner (BW) formula, we
adopt the pole positions of various resonances for parameters
appearing in the propagators.

The coupling constants appearing in relevant resonances are
determined by the empirical partial decay width of the resonances
taken from PDG~\cite{pdg2008}, and then we adjust the values of
cut-off in form factors to fit the data. The relations between the
branching ratios of the adopted resonances and the corresponding
coupling constants squared can be calculated straightforwardly with
above Lagrangians, and most of them can be found in the appendix of
Ref.~\cite{tsushima}. The detailed calculations of $g_{\rho NR}$ and
$g_{\sigma NR}$ from the $R \to N\rho(\sigma) \to N\pi\pi$ decay are
given in Ref.~\cite{xierho}. The values of coupling constants used
in our computation are compiled in the Table~\ref{coupling},
together with the properties of the resonances and the central value
of branch ratios. It should be noted that we adopt a nearly half of
the decay width of $N^*(1440)\to \Delta\pi$ in PDG as the recent
data favored~\cite{cosy,celsius,roperdecay}.

Then the invariant amplitudes can be obtained straightforwardly by
applying the Feynman rules to Fig. 1. As to the different isospin
channels, isospin coefficients are considered. We do not include
the interference terms among different diagrams because their
relative phases are not known, and the Valencia model seems to
show that such terms are very small.

\section{Numerical RESULTS AND DISCUSSION} \label{discussion}

\begin{figure}[htbp]
  \begin{center}
{\includegraphics*[scale=0.7]{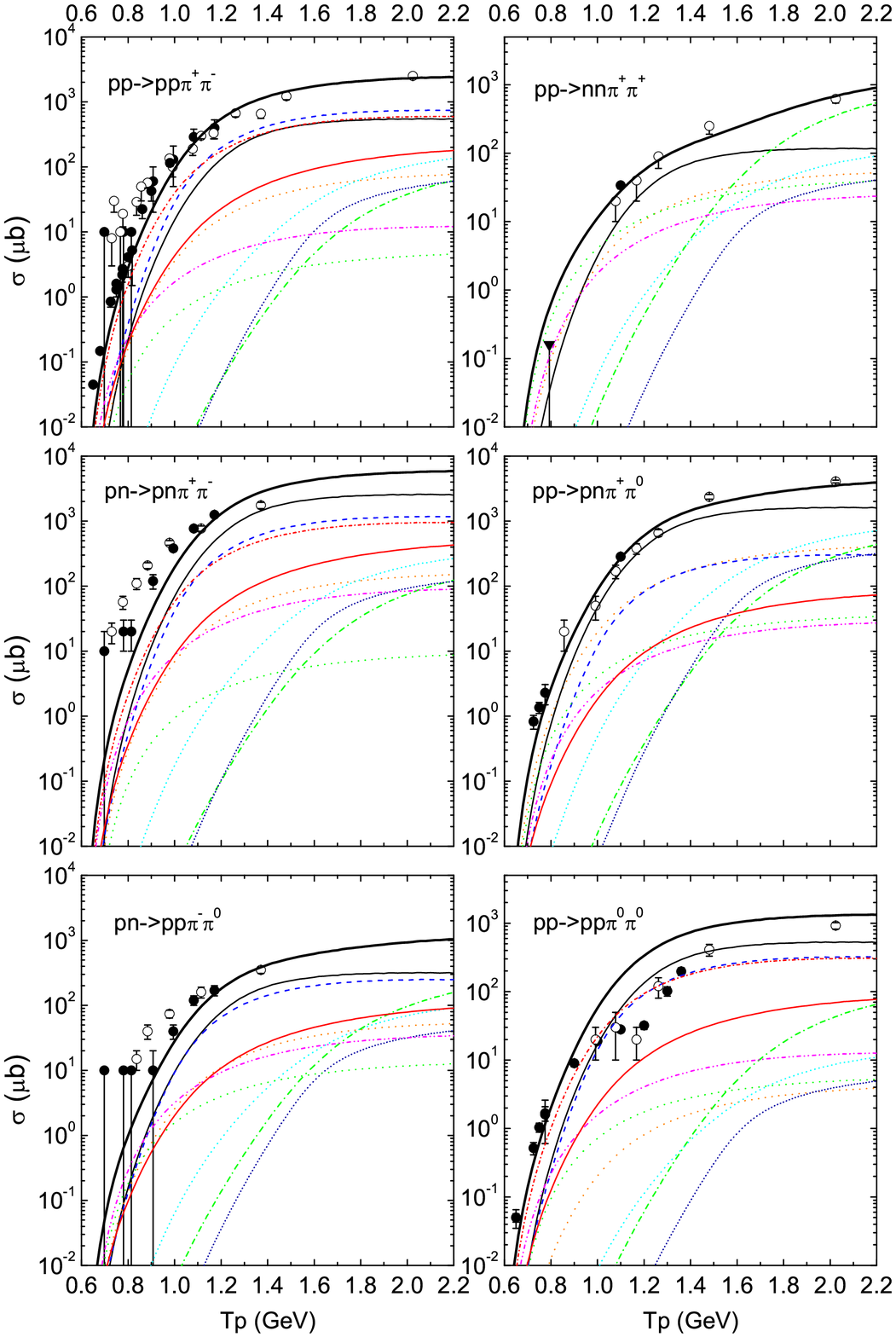}}
    \caption{(color online) Total cross sections of $NN \to NN\pi\pi$. The black solid,
    red short-dash-dotted, blue dashed, orange dotted, green dotted, cyan short-dashed,
    green dash-dotted, royal short-dotted, magenta dash-dot-dotted, pink dotted and bold
    solid curves correspond to contribution from double-$\Delta$,
    $N^*(1440) \to N\sigma$, $N^*(1440) \to \Delta\pi$, $\Delta \to \Delta\pi$,
    $\Delta \to N\pi$, $\Delta^*(1600) \to \Delta\pi$, $\Delta^*(1600) \to N^*(1440)\pi$,
    $\Delta^*(1620) \to \Delta\pi$, nucleon pole, $N \to \Delta\pi$
    and the full contributions, respectively. The solid circles and triangles represent the data from Ref.\protect\cite{cosy,celsius,otherdata,data2000}. The open circles stand for the old data\protect\cite{olddata}.} \label{tcs}
  \end{center}
\end{figure}

As a starting point, in Fig.~\ref{tcs} we demonstrate our calculated
total cross sections of six isospin channels compared with the
existing data~\cite{olddata,cosy,celsius,data2000,otherdata}.  Our
numerical results give an overall good reproduction to all six
channels. The pre-emission diagrams (see (2), (4), (6) in
Fig.~\ref{fdg}) tend to be negligibly small, consistent with the
Valencia model, so we do not include them in our concrete
computation. In Fig.~\ref{tcs} we do not show the following
negligible contributions: double-$N^*$, $N^* \to N\rho$, $N^* \to
N\pi$, $N^*(1520) \to \Delta\pi$, $N^*(1650) \to \Delta\pi$,
$N^*(1675) \to \Delta\pi$, $N^*(1680) \to \Delta\pi$, $N^*(1680) \to
N\sigma$, $N^*(1700) \to \Delta\pi$, $N^*(1710) \to \Delta\pi$,
$N^*(1710) \to N\sigma$, double-$\Delta^*(1600)$, $\Delta^*(1600)
\to N\pi$, double-$\Delta^*(1620)$, $\Delta^*(1620) \to N\rho$,
$\Delta^*(1620) \to N\pi$, double-$\Delta^*(1700)$, $\Delta^*(1700)
\to N\rho$, $\Delta^*(1700) \to N\pi$, and $\Delta^*(1700) \to
\Delta\pi$. These terms are minor either because of their small
branching ratios of double pion channel such as $N^*(1535)$,
$N^*(1650)$ and $N^*(1700)$, or belonging to higher partial waves
such as $\Delta^*(1620)$ and $N^*(1675)$, or lying beyond the
considered energies such as $N^*(1680)$, $\Delta^*(1700)$,
$N^*(1710)$ and $N^*(1720)$. It should be mentioned that
$\rho$-meson exchange is much smaller than $\pi$-meson exchange in
the available diagrams except for nucleon poles but we still include the $\rho$-meson
exchange in the calculation for the completeness of our model.

Our results underestimate the data in the close-to-threshold region
where the final state interactions (FSI) should be relevant. We do
not consider the initial state interaction (ISI) either, because at
present we do not have an unambiguous method at hand to
simultaneously include the FSI and ISI in our model. The ISI usually
has a weak energy dependence, so adjusting cut-off parameters in the
form factors may partly account for it effectively~\cite{ouyang}. We
would give some qualitative observations of FSI.

Next we shall first address the $pp \to nn\pi^+\pi^+$ channel
because it has negligible $N^*$ contribution to be more clean. Then
we shall discuss other channels and explore the different situation
at each channel. In the following we assume the same definitions of
various differential cross sections as graphically illustrated in
the experimental articles~\cite{celsius,skorodko}. The $M_{ij}$ and
$M_{ijk}$ are the invariant mass spectra, and the angular
distributions are all defined in the overall center of mass system.
The $\Theta_M$ is the scattering angle of $M$, and $\delta_{ij}$ is
the opening angle between $i$ and $j$ particles. The $\Theta^{ij}_i$ (or
$\vartheta_{i}^{ij}$ corresponding to $\widehat{\Theta}_{i}^{ij}$
defined in Ref.~\cite{celsius,skorodko}) is the scattering angle of
$i$ in the rest frame of $i$ and $j$ with respect to the beam axis
(or the sum of momenta of $i$ and $j$). The values of vertical axis
are all arbitrarily normalized.

\subsection{The channel of $pp \to nn\pi^+\pi^+$} \label{nnpipi}

In this channel, we find that the $\Delta \to N\pi\to N\pi\pi$ term
is dominant below 1000 MeV while the $\Delta \to \Delta\pi$ and
double-nucleon-pole terms are also important. The $\Delta \to
N\pi\to N\pi\pi$ term is not included in the Valencia
model~\cite{Alvarez}. Our model seems to overestimate the COSY-TOF
upper-limit by a factor of around four. The $\Delta \to \Delta\pi$
terms in two models are consistent with each other because we use
the same coupling constant of $\pi\Delta\Delta$ from quark model but
our double-$\Delta$ term contributes smaller as we use a smaller
cut-off parameter in $\pi N\Delta$ form factor. Between 1000 MeV and
1700 MeV, the contribution of the double-$\Delta$ term is the most
important one, and the $n\pi^+$ invariant mass distribution at 1100
MeV do show a clear $\Delta$ peak as can be seen in
Fig.~\ref{nnpipi}. We also find that the $\Delta \to N\pi\to
N\pi\pi$ and $\Delta \to \Delta\pi$ terms are crucial to get the
right shape of differential cross sections at 1100 MeV. Though the
data is of poor statistics, the $\pi^+\pi^+$ invariant mass spectrum
does not show obvious low-mass peak and this is realized by the
inclusion of the $\Delta \to N\pi\to N\pi\pi$ and $\Delta \to
\Delta\pi$ in our model. The $\delta_{\pi\pi}$ has also a
significant improvement compared to the double-$\Delta$ alone. These
distributions should be very useful to constrain the poorly known
coupling constant of $\pi\Delta\Delta$. The particular enhancement
compared to our model without FSI in the $nn$ invariant mass
spectrum is probably an indication of strong $^1S_0 nn$ FSI.

The contribution from $\Delta^*(1600) \to N^*(1440)\pi$ term has a
steep rise and begins to take over as the largest one for $T_p$
above 1700 MeV. Besides, at large energies contributions from the
$\Delta^*(1600) \to \Delta\pi$ and $\Delta^*(1620) \to \Delta\pi$
become significant. So in these energy region of $pp \to
nn\pi^+\pi^+$, it is a good place to explore the properties of these
$\Delta^*$ resonances. We would like to point out that these
behaviors together with the dominance of $\Delta \to N\pi\to
N\pi\pi$ and $\Delta \to \Delta\pi$ close to threshold alleviate the
isospin problem of the $pp \to nn\pi^+\pi^+$ and $pp \to
pp\pi^0\pi^0$ channels mentioned at the beginning of our article,
because the isospin coefficients of these terms in $pp \to
nn\pi^+\pi^+$ are bigger than that in $pp \to pp\pi^0\pi^0$ channel
and this is contrary to the case of double-$\Delta$. As a result, we
get an improvement on the description of all isospin channels.

It should be addressed that it is very useful to pin down the
cut-off values in form factors of the relevant $\Delta$ and
$\Delta^*$ contributions using the data of $pp \to nn\pi^+\pi^+$ at
first, and then it makes much easier for us to determined the $N^*$
contributions in other channels. The new value of total cross
section of $pp \to nn\pi^+\pi^+$ measured at CELSIUS~\cite{celsius}
is in line with previous data and this gives our some confidence on
the extracted parameters. Further accurate measurements of the $pp
\to nn\pi^+\pi^+$ channel should be very helpful for the improvement
of the model.

\subsection{The channel of $pp \to pp\pi^+\pi^-$} \label{2picharge}

Below 1000 MeV the $N^*(1440) \to N\sigma$ term is the largest while
the $N^*(1440) \to \Delta\pi$ term is the second. Of these two terms
the $\sigma$-meson exchange gives much bigger contribution than the
$\pi$-meson exchange as depicted in Fig.~\ref{nstar1440} and this
shows the importance of isoscalar excitation of $N^*(1440)$. The
Double-$\Delta$ term is negligible at this low energies as well as
the $\Delta \to \Delta\pi$ term. Contributions from the nucleon pole
and $N \to \Delta\pi$ terms are visible below 700 MeV. The proton
and pion angular distributions in the center of mass system at 650
and 680 MeV are trivially isotropic and the model does agree with
the measured data~\cite{celsius}. So we do not show them here. The
differential cross sections at 750, 775, 800 and 895 MeV are given
in Fig.~\ref{pp2pi750}, Fig.~\ref{pp2pi775}, Fig.~\ref{pp2pi800} and
Fig.~\ref{pp2pi895}, respectively. Our model calculations reproduce
the published data well and are also compatible to the very
preliminary data of CELSIUS at 895 MeV~\cite{skorodko} as shown in
Fig.~\ref{pp2pi895}. The role of $N^*(1440) \to \Delta\pi$ is
evident in invariant mass spectrums and the data is fitted better
than that including $N^*(1440) \to N\sigma$ alone. Most obviously,
the anisotropic shape of $\vartheta_{\pi^+}^{\pi\pi}$ is well fitted
after including the $N^*(1440) \to \Delta\pi$ while $N^*(1440) \to
N\sigma$ term is symmetric, so $\vartheta_{\pi^+}^{\pi\pi}$ together
with $\vartheta_{\pi^+}^{p\pi^+}$ and $\vartheta_{\pi^-}^{p\pi^-}$
is used to determined the ratio of partial decay widths of
$N^*(1440) \to \Delta\pi$ and $N^*(1440) \to
N\sigma$~\cite{celsius}. The results are strongly energy dependent
and give a smaller decay width of $N^*(1440) \to \Delta\pi$ than
that listed in PDG. This is believed to support the breathing mode
of $N^*(1440)$~\cite{ropertheory}. The FSI is evident in $pp$
invariant mass spectrum, but seems to be much weaker than in the
case of $pp \to nn\pi^+\pi^+$ channel.

In the Valencia model the double-$\Delta$ is dominant above 1300
MeV. However, because we use smaller cut-off parameter for the $\pi
N\Delta$ form factor in order to fit both $nn\pi^+\pi^+$ and
$pp\pi^0\pi^0$ channels, our model shows that $N^*(1440) \to
\Delta\pi$ begins to take over above 1100 MeV, and double-$\Delta$
and $N^*(1440) \to N\sigma$ are also important and comparable. We
give the differential cross sections at 1100MeV and 1360 MeV in
Fig.~\ref{pp2pi1100} and Fig.~\ref{pp2pi1360} which can be tested by
the measured data of CELSIUS~\cite{skorodko}. The prominent features
are the double hump structure in $M_{\pi^+\pi^-}$ and the upward
bend in $\delta_{\pi^+\pi^-}$ which arise from the $N^*(1440) \to
\Delta\pi$. The Valencia model give very similar results  because
$M_{\pi^+\pi^-}$ and $\delta_{\pi^+\pi^-}$ are sensitive to the
appearance of $N^*(1440) \to \Delta\pi$. These seem to somewhat
incompatible to the preliminary data~\cite{dataIJMPA,skorodko} which
show the phase space behavior in these two spectrums. The same
phenomena also happen in the channel of $pp \to pp\pi^0\pi^0$ at
high energies, and we will discuss them altogether later.

\subsection{The channel of $pp \to pp\pi^0\pi^0$} \label{2pineutron}

The $N^*(1440) \to N\sigma$ term dominates below 1000 MeV and the
nucleon pole term also gives significant contribution below 800 MeV.
The $N^*(1440) \to \Delta\pi$ and double-$\Delta$ contributions are
comparable in this energy region. So it should be cautious to use
this channel to extract the ratio of the partial decay widths for
the decay of $N^*(1440)$. Indeed, the extracted ratios from $pp \to
pp\pi^0\pi^0$ are about one third of those from $pp \to
pp\pi^+\pi^-$ at the same nominal mass under the assumption that
$N^*(1440)$ dominates in this energy range~\cite{celsius,skorodko}.
The significant double-$\Delta$ and nucleon pole contributions might
account for this discrepancy and should be reasonably incorporated
in the fit.

Above 1100~MeV, the double-$\Delta$ term dominates; the $N^*(1440)
\to \Delta\pi$ and $N^*(1440) \to N\sigma$ are also important and
give similar contributions. Other contributions are much smaller.
The most striking feature in this energy region is that a level-off
behavior happens in the total cross section between 1000 and 1200
MeV, while other channels rise smoothly when increasing the incident
energy. Our model fails to describe this behavior and also
overestimates the high energy data. It is possible that this shape
is caused by the interference of different diagrams which are not
included in our model, but this would require a peculiar energy
dependence of $N^*$ as shown by the isospin
decomposition~\cite{celsius,skorodko}. Another possible explanation
is that there maybe exist a steep rise of some kind of contribution
when other contributions are saturated in this energy region. This
happens in the channel of $pp \to nn\pi^+\pi^+$ where a weak
level-off at 1600 MeV is caused by the steep rise of $\Delta^*(1600)
\to N^*(1440)\pi$. However, this is not the case for the $pp \to
pp\pi^0\pi^0$ channel where $\Delta^*(1600) \to N^*(1440)\pi$ gives
much smaller contribution due to the isospin factor. So this problem
is left for further clarification.

In Fig.~\ref{2pi0775}, Fig.~\ref{2pi0895} and Fig.~\ref{2pi01000},
we show the differential cross sections of $pp \to pp\pi^0\pi^0$ at
beam energies of 775, 895 and 1000 MeV, respectively. The data at
775 MeV are well reproduced and $N^*(1440) \to N\sigma$ is
overwhelmingly dominant. Some of the angular distributions are
sensitive to the presence of the $N^*(1440) \to \Delta\pi$
contribution, and hence can be used to determine the partial decay
ratios of $N^*(1440)$, although this is somewhat complicated by the
double-$\Delta$ and nucleon pole contributions as we pointed out
earlier. The contribution of $N^*(1440) \to \Delta\pi$ and
double-$\Delta$ terms become much clearer at 895 and 1000 MeV,
though slight discrepancy in invariant mass spectrums at 895 MeV
exists between our model and the measured data, which remind us that
it needs a further improvement in the crossover region. The phase
space shapes of $M_{\pi^0\pi^0}$ and $\delta_{\pi^0\pi^0}$ begin to
appear ever since 895 MeV and up to high energies of this channel as
shown in Fig.~\ref{2pi01100}, Fig.~\ref{2pi01200} and
Fig.~\ref{2pi01300} which are the differential cross sections of $pp
\to pp\pi^0\pi^0$ at 1100, 1200 and 1300 MeV. However, because the
influence of the $N^*(1440) \to \Delta\pi$ does not decrease much or
disappear at these energies, our model gives a double hump structure
in $M_{\pi^0\pi^0}$ and forward peak in $\delta_{\pi^0\pi^0}$ which
are contradictory to the CELSIUS data. In order to explain the data,
it is required that the $N^*(1440) \to \Delta\pi$ shows up at low
energies but immediately saturated at about 1000 MeV. That is a
peculiar energy dependence behavior which does not supported by our
model. Except for $M_{\pi\pi}$ and $\delta_{\pi\pi}$, other
spectrums at high energies are well fitted by our model both in $pp
\to pp\pi^+\pi^-$ and $pp \to pp\pi^0\pi^0$. So we would rather
conclude that something happens in the $\pi\pi$ system which needs a
more thorough investigation as the next step. The $\pi$-$\pi$
rescattering is found to be negligible at these
energies~\cite{piscatter}.

The effect of FSI is not obvious at low energies compared to our
calculated curve but enhancement seems to happen at high energies.
It is possible that this is related to the behavior of $\pi^0\pi^0$
system.

\subsection{The channels of $pp \to pn\pi^+\pi^0$, $pn \to pp\pi^-\pi^0$
and $pn \to pn\pi^+\pi^-$\label{pntopp}}

The $N^*(1440) \to N\sigma$ does not present in the $pp \to
pn\pi^+\pi^0$ reaction, so the double-$\Delta$ term is the most
important one in a wide energy range. The $\Delta \to \Delta\pi$ and
$\Delta \to N\pi\to N\pi\pi$ terms have significant contribution
below 800 MeV and also have some contribution at higher energies
together with the $\Delta^*(1600)$ and $\Delta^*(1620)$ terms. The
agreement with the data is very good and the FSI may influence the
near-threshold region since our model slightly underestimates this
part.

The channel of $pn \to pp\pi^-\pi^0$ is another reaction where the
$N^*(1440) \to N\sigma$ does not contribute. Since the charged meson
exchange is allowed in this channel, the $N^*(1440) \to \Delta\pi$
term is very important and is of the same order as the
double-$\Delta$ term in the whole energy region. The contributions
from the nucleon pole and $\Delta \to N\pi \to NN\pi$ terms are also quite
significant near the threshold. Our results reproduce the new bubble
chamber data measured by KEK~\cite{data2000} very well, but
underestimate the old data~\cite{olddata} by about a factor of 5.
Since the double-$\Delta$ contribution has been well determined by
the channel of $pp \to nn\pi^+\pi^+$, we think that the main
ambiguity comes from the $N^*(1440) \to \Delta\pi$ term. If future
experiments confirm the old data, then the isovector mesons like
$\pi$- and $\rho$-meson should play more important role in the
excitation of $N^*(1440)$. On the other hand, the new data of KEK
support the isoscalar excitation of $N^*(1440)$ which is favored by
our model.

The $pn \to pn\pi^+\pi^-$ channel is interesting because it can shed
light on the low mass enhancement in $M_{\pi\pi}$, known as the ABC
effect of double pion production in nuclear fusion
reactions~\cite{abceffect}. Below 900 MeV, the $N^*(1440) \to
N\sigma$ is found to be dominant while the double-$\Delta$ and
$N^*(1440) \to \Delta\pi$ terms also give some contribution. The
nucleon pole and $N \to \Delta\pi$ terms are also important close to
threshold. Above 1000 MeV, the double-$\Delta$ term is the most
important one and the $N^*(1440)$ gives sizable contribution at high
energies. The total contribution gives a reasonable description to
the new KEK data while the underestimation of the data close to
threshold may be due to the omission of the $pn$ FSI. Similar to the
$pn \to pp\pi^-\pi^0$ channel, our model does not favor the old
bubble chamber data which need large isovector excitation of
$N^*(1440)$. Very recently the ABC effect is experimentally
established in $pn \to d\pi^0\pi^0$ at beam energies of 1.03 and
1.35 GeV, and has been interpreted as an s-channel double-$\Delta$
resonance~\cite{abceffect}. According to the observation of our
model, the $N^*(1440)$ emerges at these energies so it is necessary
to take a further look at the mechanism of ABC effect in $pn \to
d\pi^0\pi^0$ reaction. As a matter of fact, it has been demonstrated
that at beam momentum of 1.46 GeV (corresponding to beam energies of
800 MeV) where the $N^*(1440)$ is expected to be dominant, the
deuteron momentum spectra can be reasonably explained by the
interference of the $N^*(1440) \to N\sigma$ and $N^*(1440) \to
\Delta\pi$~\cite{Alvarez,Alvarezabc}.

\subsection{Final State Interaction}

As discussed above, the effect of FSI is anticipated to influence
the results close to threshold where the s-wave is expected to be
dominant. Usually the Jost function is used to account for the FSI
enhancement factor,
\begin{equation}
 {J(k)}^{ - 1}  = \frac{{k + i\beta }}{{k - i\alpha }},
\end{equation}
where $k$ is the relative momentum of $NN$ subsystem in the final
state. The corresponding scattering length and effective range are:
\begin{equation}
a = \frac{{\alpha  + \beta }}{{\alpha \beta }} ,
\begin{array}{*{20}c}
   {} & {}  \\
\end{array}
r =\frac{2}{{\alpha  + \beta }} \label{eq:fsia} ,
\end{equation}
with $a$ = -7.82fm and $r$ = 2.79fm for $^{1}S_{0}$ $pp$
interaction, $a$ = 5.42fm and $r$ = 1.76fm for $^{3}S_{1}$ isoscalar
$pn$ interaction, and $a$ = -18.45fm and $r$ = 2.83fm for
$^{1}S_{0}$ $nn$ interaction. At higher energies, the high partial
waves become important and above approximate treatment would
deteriorate rapidly. Fortunately, the effect of FSI should
significantly decrease. So we may just ignore it above 1.4 GeV. To
investigate the influence of the FSI to the energy dependence of
cross sections, we assume the Jost function for the FSI and
normalize this factor to the unity at the beam energy of 1.4GeV. For
the final states with the $pn$ pair, we assume it is mainly in the
$^{3}S_{1}$ isoscalar state.

Though above prescription is quite rough, the agreement with the
data are considerably improved. In Fig.~\ref{tcsfsi}, we demonstrate
the total cross section below 1.4GeV. The effect of FSI can be seen
in some of the differential cross sections especially $NN$
spectrums. In Fig.~\ref{nnpipif} we take $pp \to nn\pi^+\pi^+$
channel as a typical example. The $nn$ FSI gives a sharp peak in the
$nn$ spectrums which agree with the $pp \to nn\pi^+\pi^+$ data.
However, the data in other channels do not favor this sharp peak and
this reflects the drawbacks of our formalism. The $nn$ FSI slightly
improves the fit of the $n\pi^+\pi^+$ and $nn\pi^+$ spectrums but
increase the slope of $\delta_{n\pi^+}$. FSI has very small
influence on other angular distributions. The situations for other
channels are similar.

\section{Summary} \label{summary}

In this article, we present a simultaneous analysis of varous
isospin channels of double pion production in nucleon-nucleon
collisions up to 2.2 GeV within an effective Lagrangian approach. We
study the contributions of various resonances with mass up to
1.72~GeV and demonstrate that $N^*(1440)$, $\Delta$,
$\Delta^*(1600)$, $\Delta^*(1620)$ and nucleon pole constitute the
main ingredients to reasonably explain the measured data while the
contribution of other resonances are negligible. We suggest that it
is necessary to consider other influence such as the double-$\Delta$
and nucleon pole contributions when one studies the properties of
$N^*(1440)$ in the channels of $pp \to pp\pi^+\pi^-$ and $pp \to
pp\pi^0\pi^0$. Our model well describe the measured differential
cross sections except some $\pi\pi$ spectra which are left as an
open problem. Compared with the Valencia model, the main differences
are: (1) Among 3 major ingredients, double-$\Delta$,
$N^*(1440)\to\Delta\pi$ and $N^*(1440)\to N\sigma$ terms, considered
in the Valencia model, our model increases significantly the
relative contribution from the $N^*(1440)\to N\sigma$ term by
reducing the relative branching ratio of $N^*(1440)\to\Delta\pi$ and
assuming a smaller cut-off parameter for the $\pi N\Delta$ coupling;
(2) In addition, our model introduces significant contributions from
$\Delta\to N\pi\to N\pi\pi$ at energies near threshold and from
$\Delta^*(1600)$ and $\Delta^*(1620)$ at energies above 1.5 GeV.
Though the model should be improved to reasonably incorporated the
ISI and FSI, the conclusions reached from our model should be
helpful to the future experiments to be performed at COSY and
HIRFL-CSR as well as further theoretical study on related problems.
Our results also give hints to the ABC effect in the $pn \to
d\pi^0\pi^0$ and $pd \to ^3$He$\pi^0\pi^0$ reactions which need to
be further explored.

\begin{acknowledgments}

Useful discussions with E. Oset, J. J. Xie and Z. Ouyang are
gratefully acknowledged. We also thank T. Skorodko, E.
Dorochkevitch, H. Clement and B. H\"{o}istad for providing the data
files. Special thanks to C. Wilkin for careful reading of the
manuscript and stimulating comments. This work was supported by the
National Natural Science Foundation of China (Nos. 10635080,
10875133, 10821063, 10925526 0701180GJ0).

\end{acknowledgments}

\begin{figure}[htbp]
  \begin{center}
{\includegraphics*[scale=1.0]{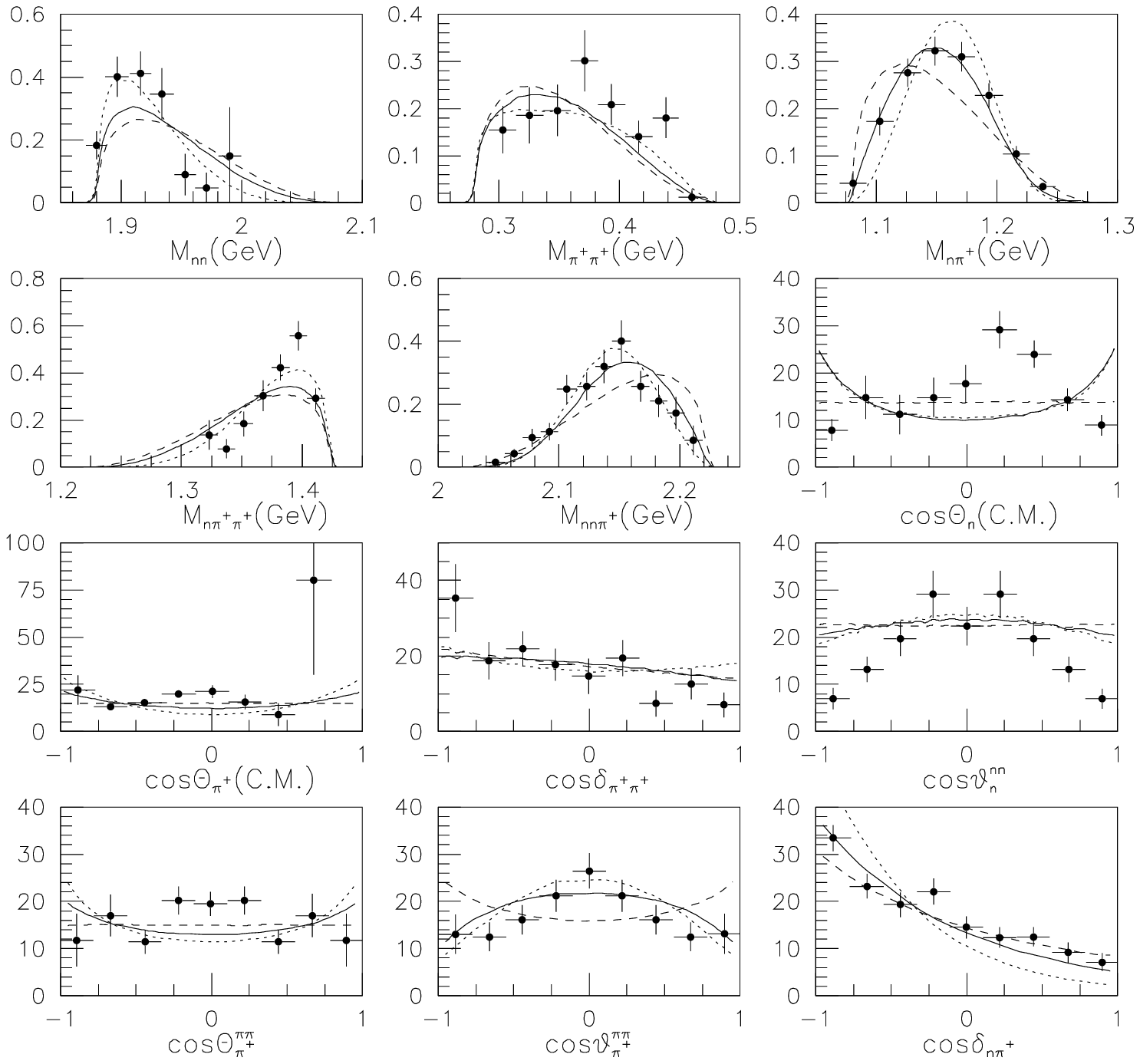}}
    \caption{Differential cross sections of $pp \to nn\pi^+\pi^+$
    at beam energies 1100 MeV. The dashed, dotted and solid curves
    correspond to the phase space, double- $\Delta$ and full model
    distributions, respectively. The data are from
    Ref.~\protect\cite{dataIJMPA}.} \label{nnpipi}
  \end{center}
\end{figure}
\begin{figure}[htbp]
  \begin{center}
{\includegraphics*[scale=0.8]{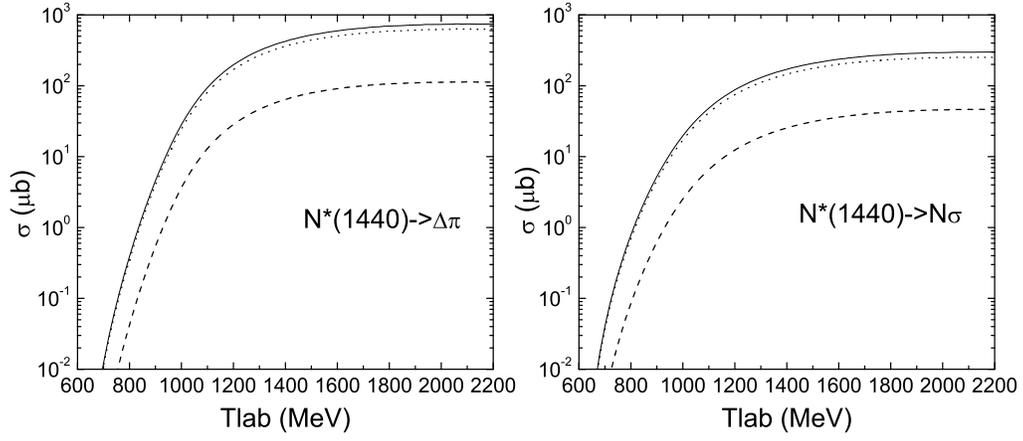}}
    \caption{The $N^*(1440) \to \Delta\pi$, $N^*(1440) \to N\sigma$
    terms of $pp \to pp\pi^+\pi^-$. The dashed, dotted and solid curves
    correspond to $\pi$-meson exchange, $\sigma$-meson exchange and
    total contribution.} \label{nstar1440}
  \end{center}
\end{figure}
\begin{figure}[htbp]
  \begin{center}
{\includegraphics*[scale=1.0]{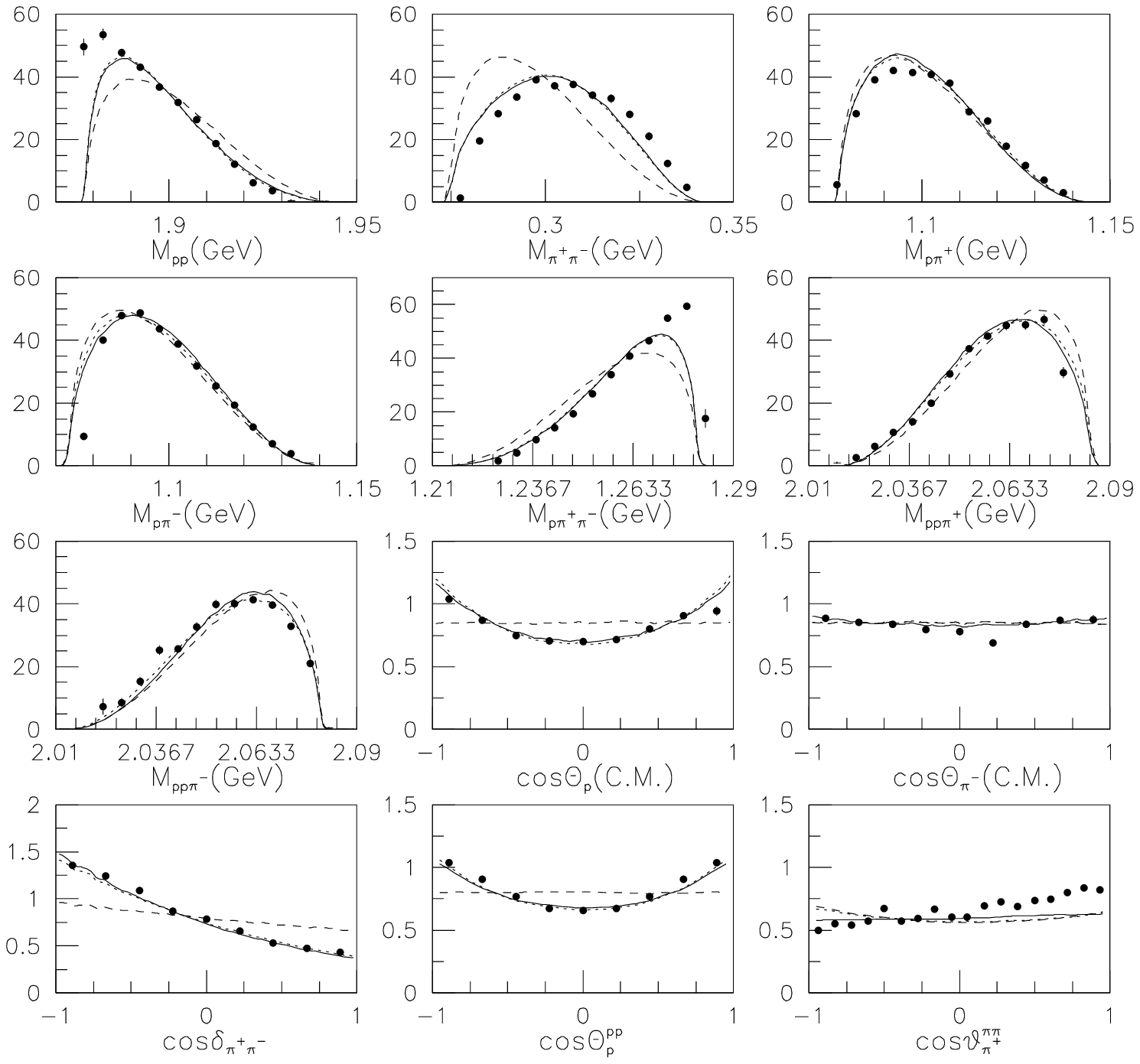}}
    \caption{Differential cross sections of $pp \to pp\pi^+\pi^-$
    at beam energies 750 MeV. The dashed, dotted and solid curves
    correspond to the phase space, $N^*(1440)\to N\sigma$ and full
    model distributions, respectively. The data are from
    Ref.~\protect\cite{celsius}.} \label{pp2pi750}
  \end{center}
\end{figure}
\begin{figure}[htbp]
  \begin{center}
{\includegraphics*[scale=1.0]{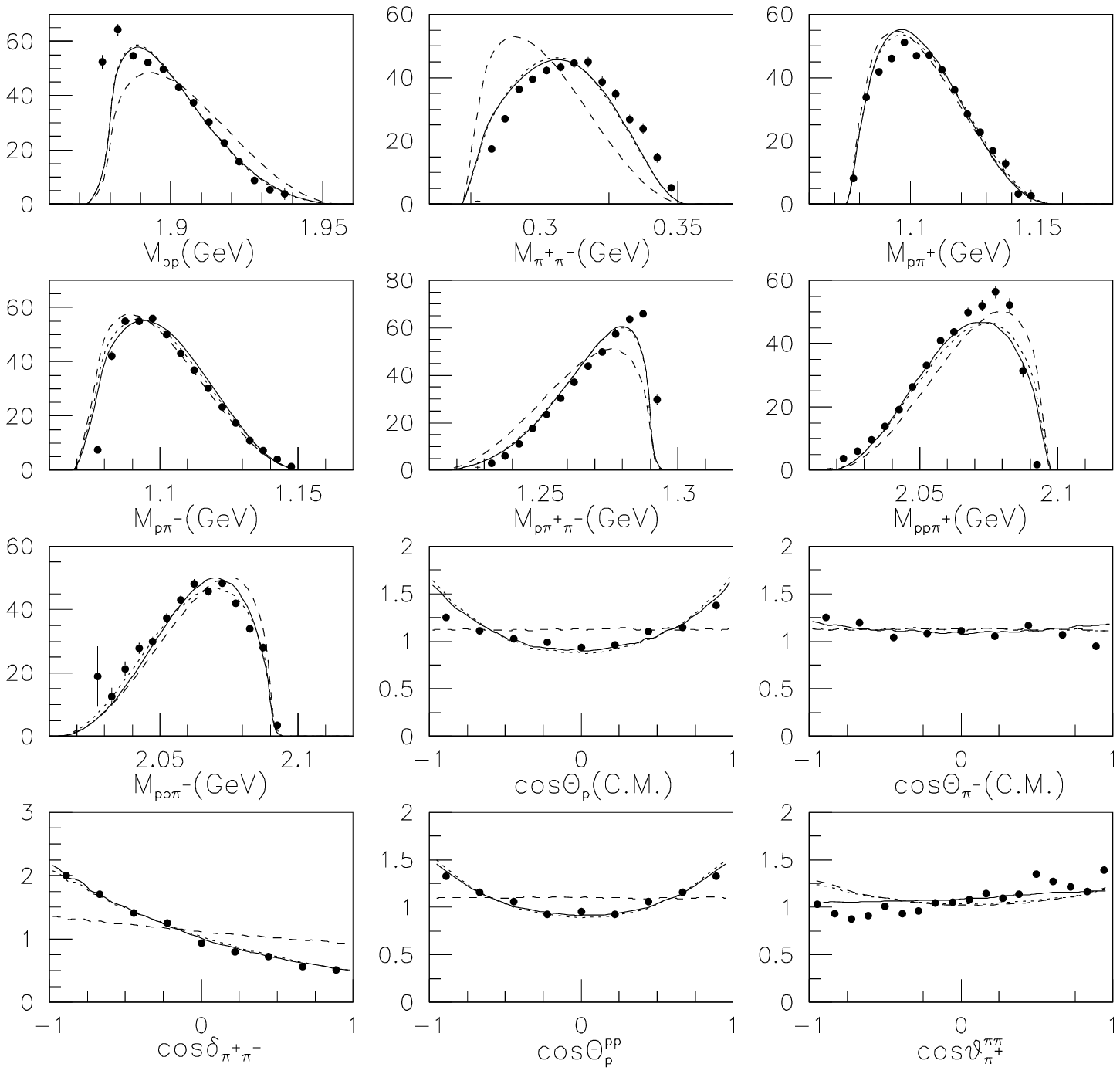}}
    \caption{Differential cross sections of $pp \to pp\pi^+\pi^-$
    at beam energies 775 MeV.  The meaning of curves is the same
    as Fig.~\protect\ref{pp2pi750}. The data are from
    Ref.~\protect\cite{celsius}.} \label{pp2pi775}
  \end{center}
\end{figure}
\begin{figure}[htbp]
  \begin{center}
{\includegraphics*[scale=1.0]{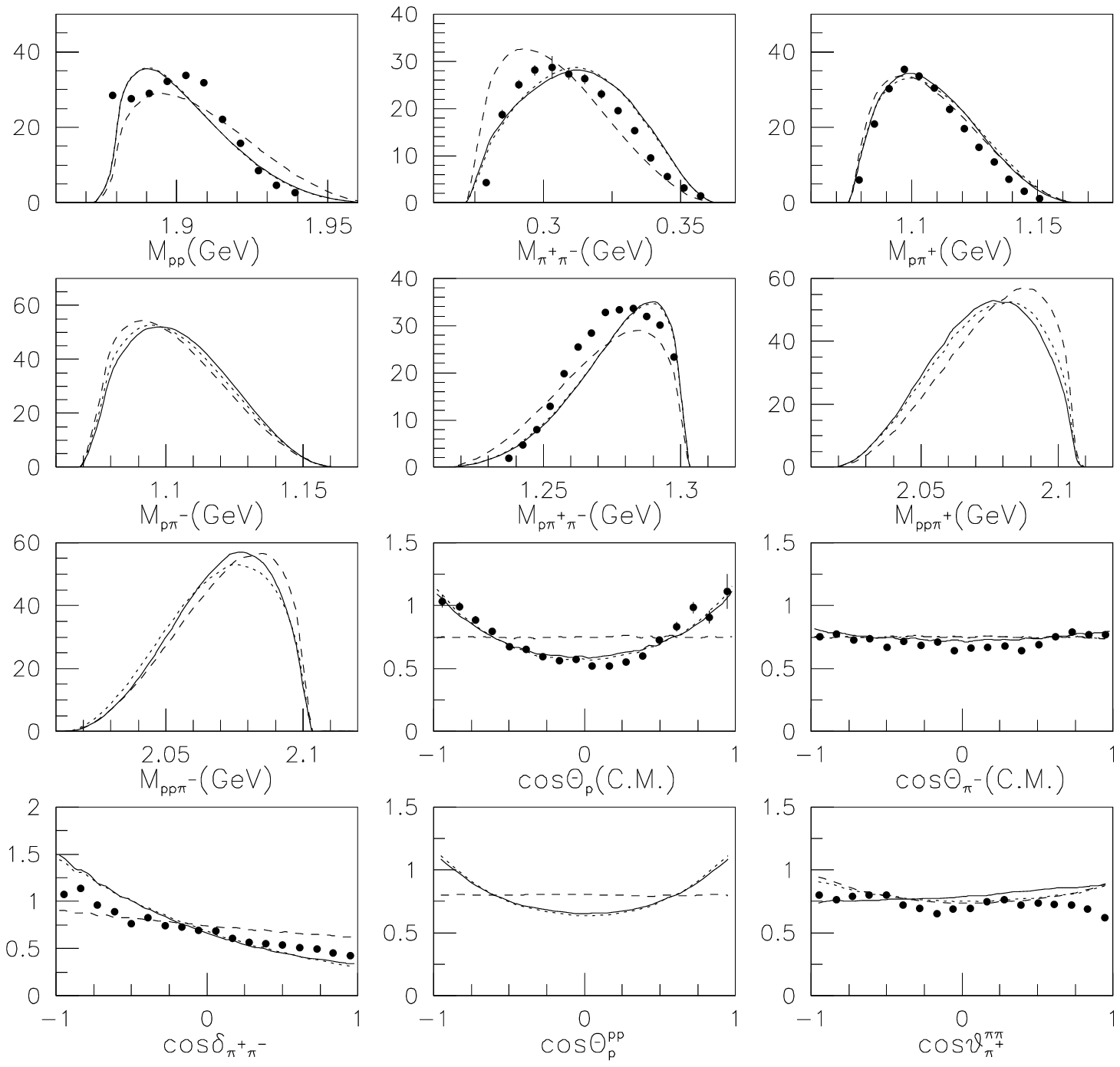}}
    \caption{Differential cross sections of $pp \to pp\pi^+\pi^-$
    at beam energies 800 MeV. The meaning of curves is the same as
    Fig.~\protect\ref{pp2pi750}. The data are from
    Ref.~\protect\cite{cosy}.} \label{pp2pi800}
  \end{center}
\end{figure}
\begin{figure}[htbp]
  \begin{center}
{\includegraphics*[scale=0.5]{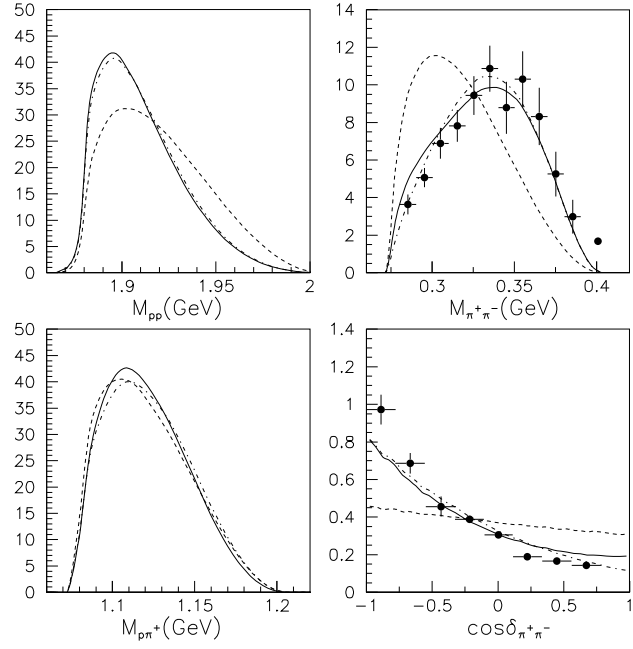}}
    \caption{Differential cross sections of $pp \to pp\pi^+\pi^-$
    at beam energies 895 MeV. The meaning of curves is the same as
    Fig.~\protect\ref{pp2pi750}. The preliminary data are from
    Ref.~\protect\cite{skorodko}.} \label{pp2pi895}
  \end{center}
\end{figure}
\begin{figure}[htbp]
  \begin{center}
{\includegraphics*[scale=0.5]{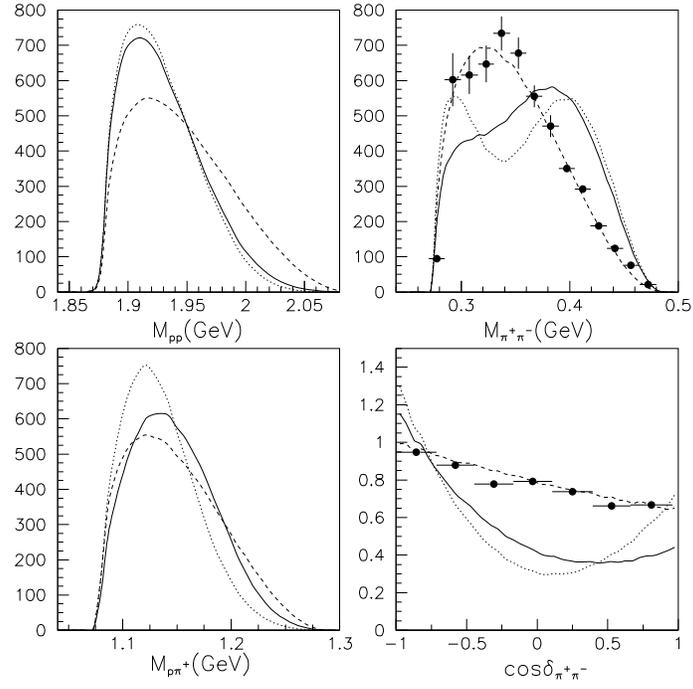}}
    \caption{Differential cross sections of $pp \to pp\pi^+\pi^-$
    at beam energies 1100 MeV. The dashed, dotted and solid curves
    correspond to the phase space, $N^*(1440)\to \Delta\pi$ and full
    model distributions, respectively. The preliminary data are from
    Ref.~\protect\cite{dataIJMPA}.} \label{pp2pi1100}
  \end{center}
\end{figure}
\begin{figure}[htbp]
  \begin{center}
{\includegraphics*[scale=0.5]{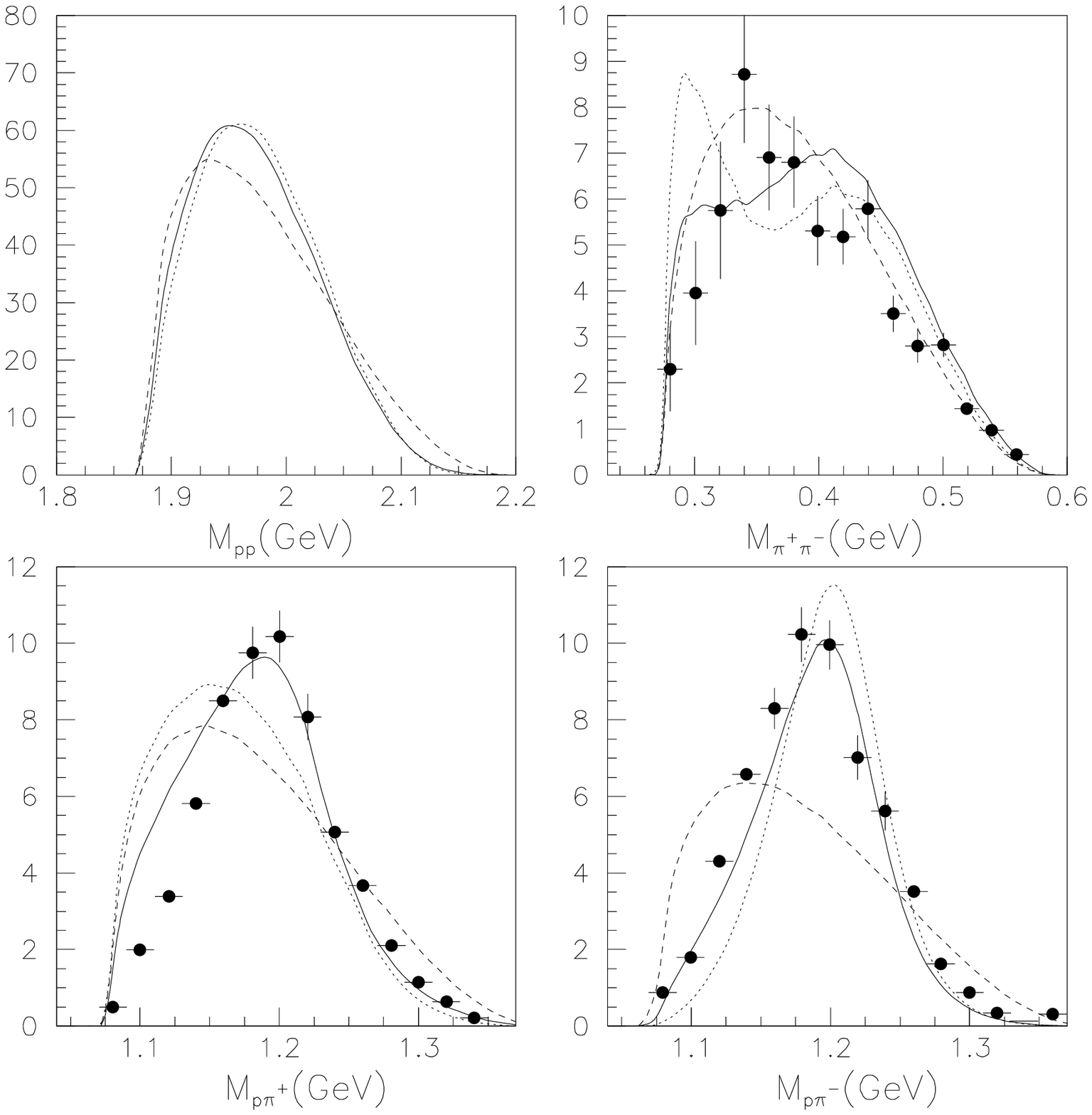}}
    \caption{Differential cross sections of $pp \to pp\pi^+\pi^-$
    at beam energies 1360 MeV. The meaning of curves is the same
    as Fig.~\protect\ref{pp2pi1100}. The data are from
    Ref.~\protect\cite{skorodko}.} \label{pp2pi1360}
  \end{center}
\end{figure}
\begin{figure}[htbp]
  \begin{center}
{\includegraphics*[scale=1.0]{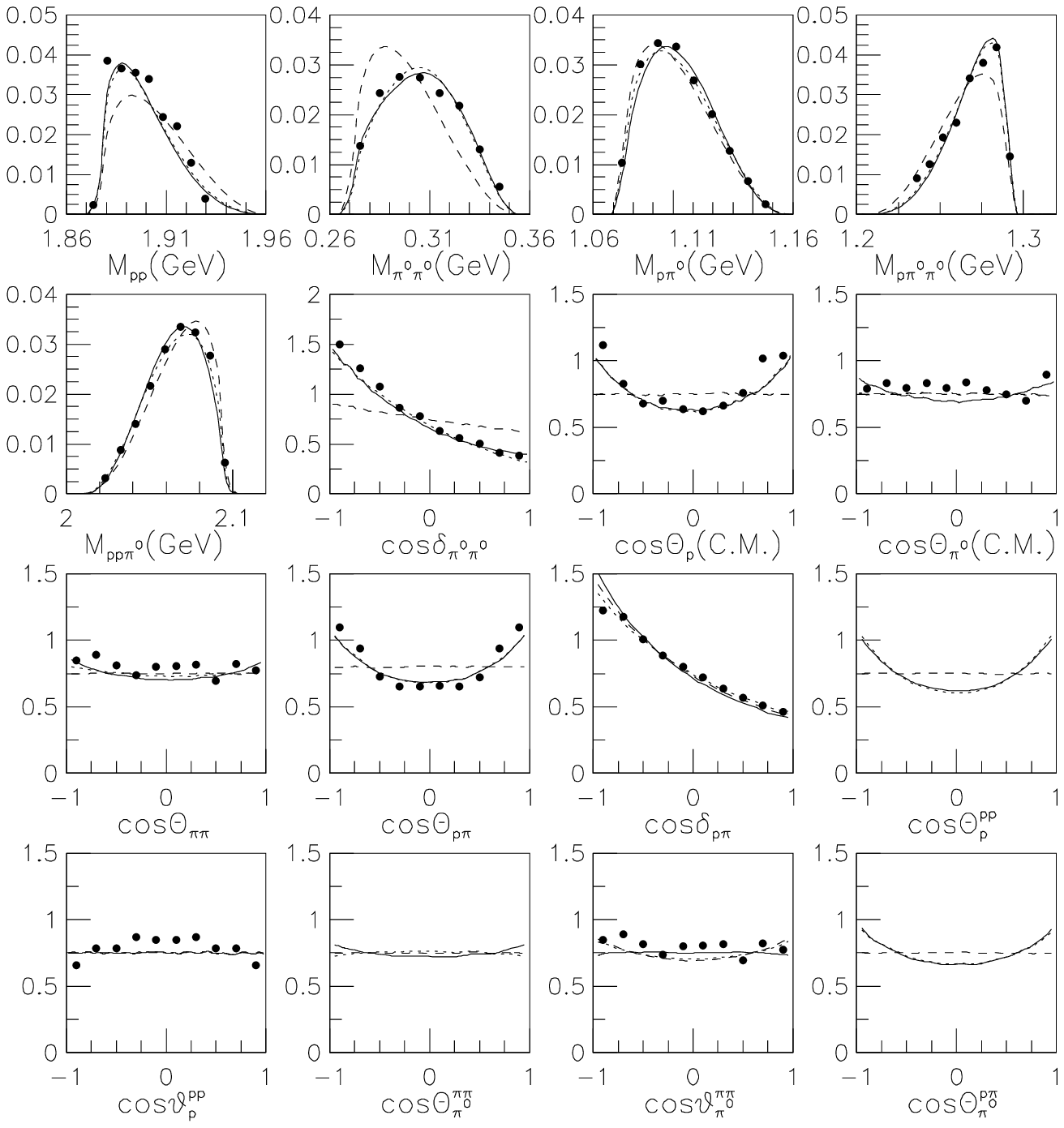}}
    \caption{Differential cross sections of $pp \to pp\pi^0\pi^0$
    at beam energies 775 MeV.  The meaning of curves is the same
    as Fig.~\protect\ref{pp2pi750}. The data are from
    Ref.~\protect\cite{skorodko}.} \label{2pi0775}
  \end{center}
\end{figure}
\begin{figure}[htbp]
  \begin{center}
{\includegraphics*[scale=1.0]{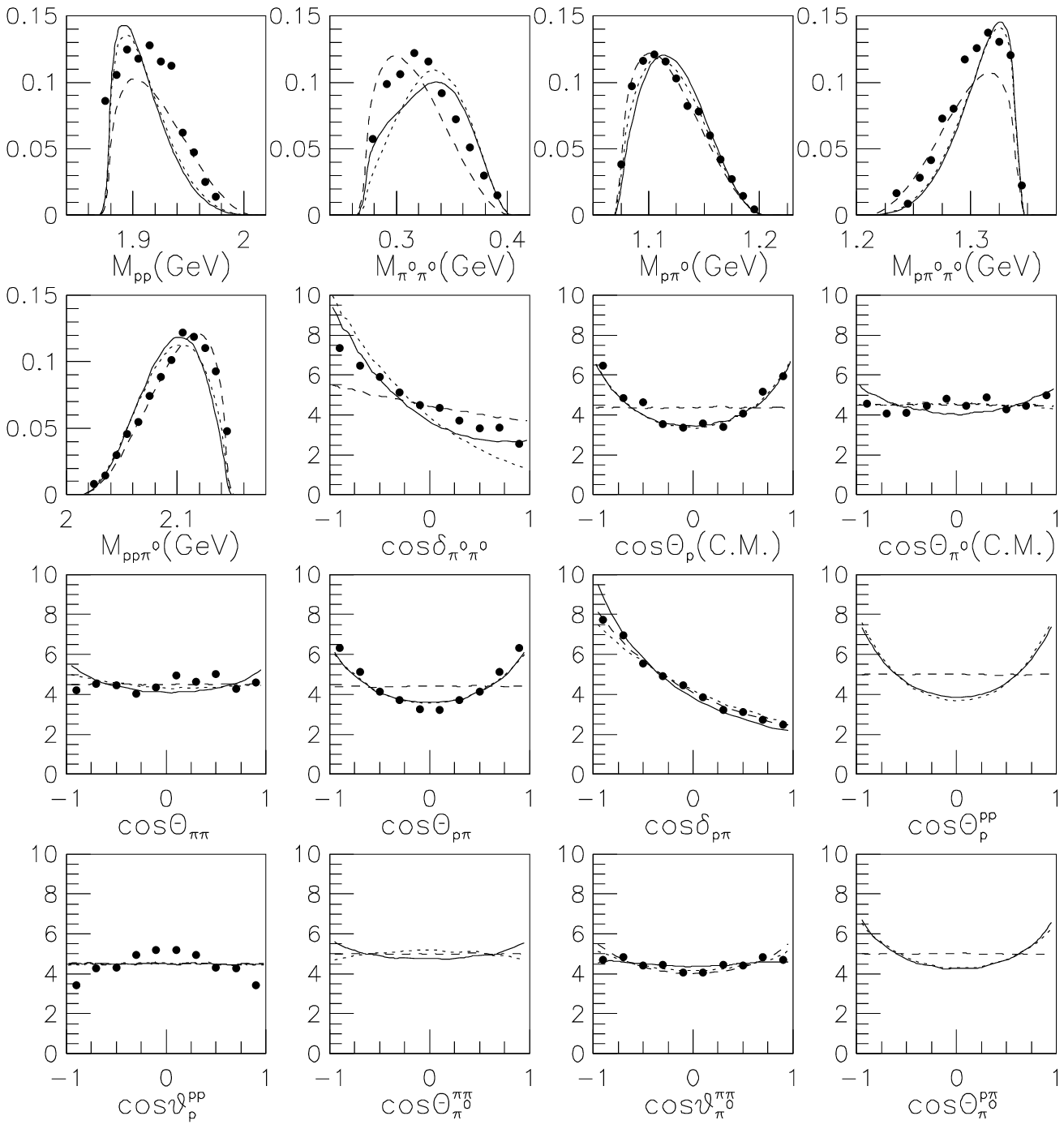}}
    \caption{Differential cross sections of $pp \to pp\pi^0\pi^0$
    at beam energies 895 MeV. The meaning of curves is the same as
    Fig.~\protect\ref{pp2pi750}. The data are from
    Ref.~\protect\cite{skorodko}.} \label{2pi0895}
  \end{center}
\end{figure}
\begin{figure}[htbp]
  \begin{center}
{\includegraphics*[scale=1.0]{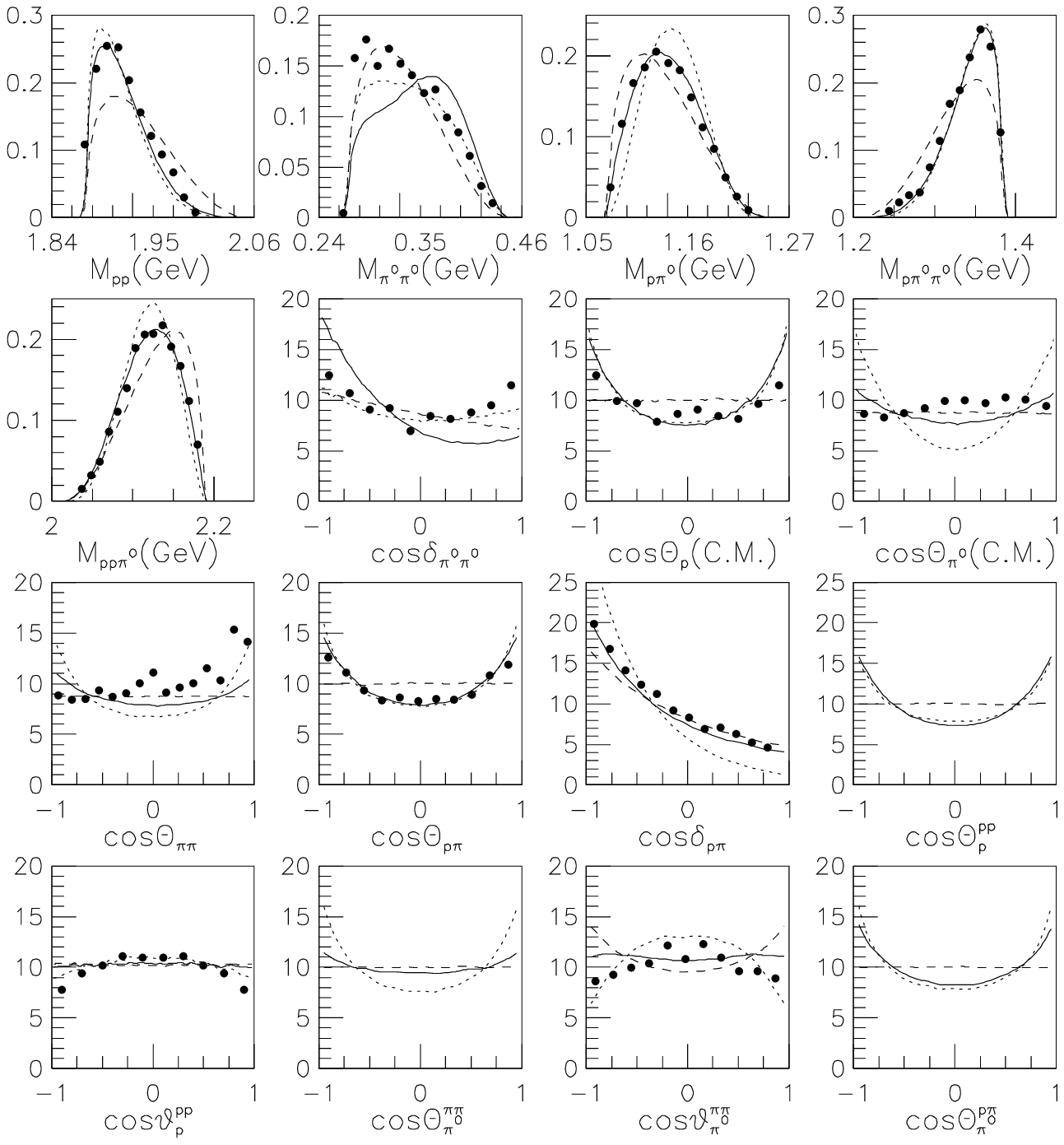}}
    \caption{Differential cross sections of $pp \to pp\pi^0\pi^0$
    at beam energies 1000 MeV. The dashed, dotted and solid curves
    correspond to the phase space, double- $\Delta$ and full model
    distributions, respectively. The data are from
    Ref.~\protect\cite{skorodko}.} \label{2pi01000}
  \end{center}
\end{figure}
\begin{figure}[htbp]
  \begin{center}
{\includegraphics*[scale=1.0]{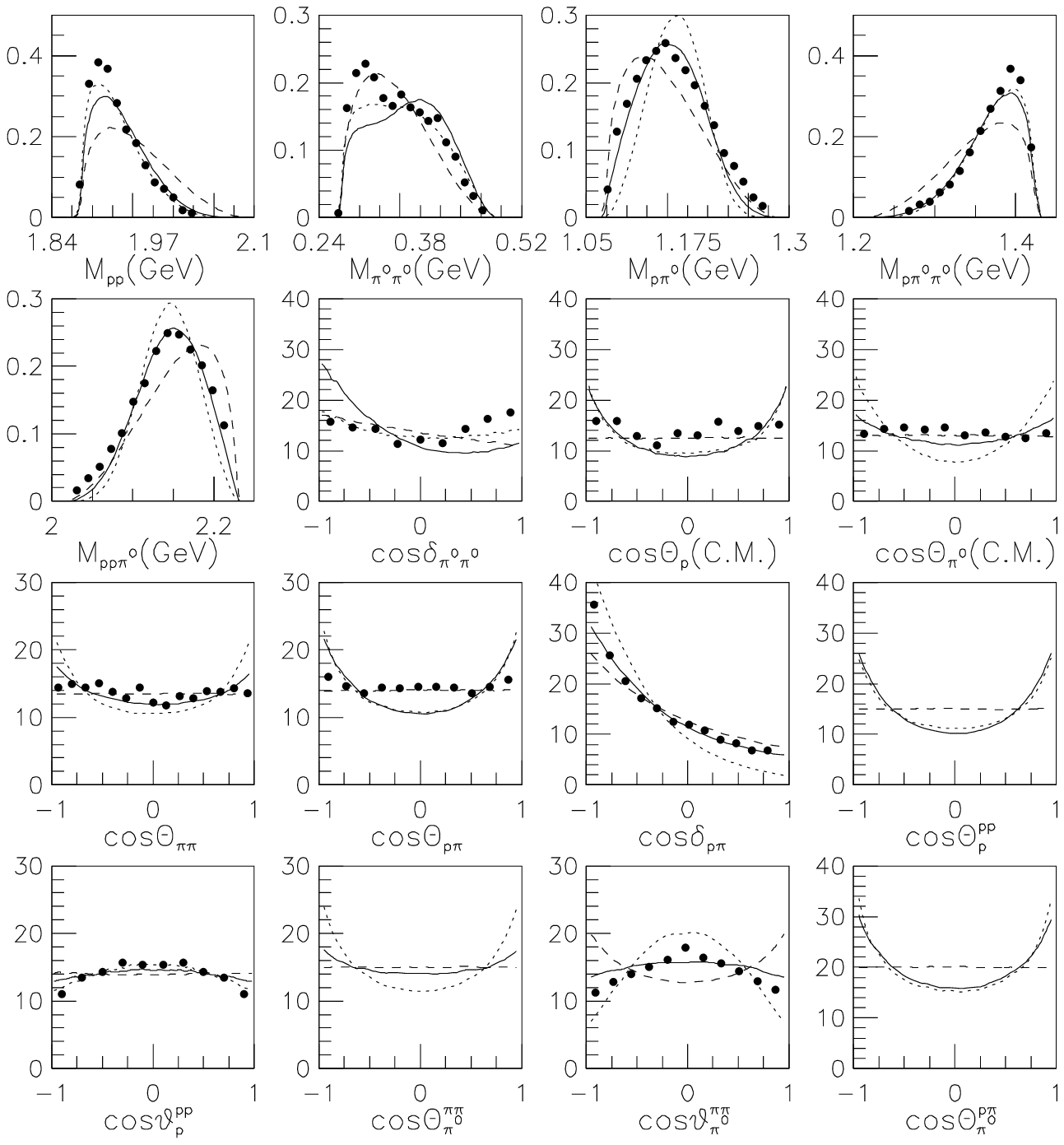}}
    \caption{Differential cross sections of $pp \to pp\pi^0\pi^0$
    at beam energies 1100 MeV. The meaning of curves is the same as
    Fig.~\protect\ref{2pi01000}. The data are from
    Ref.~\protect\cite{skorodko}.} \label{2pi01100}
  \end{center}
\end{figure}
\begin{figure}[htbp]
  \begin{center}
{\includegraphics*[scale=1.0]{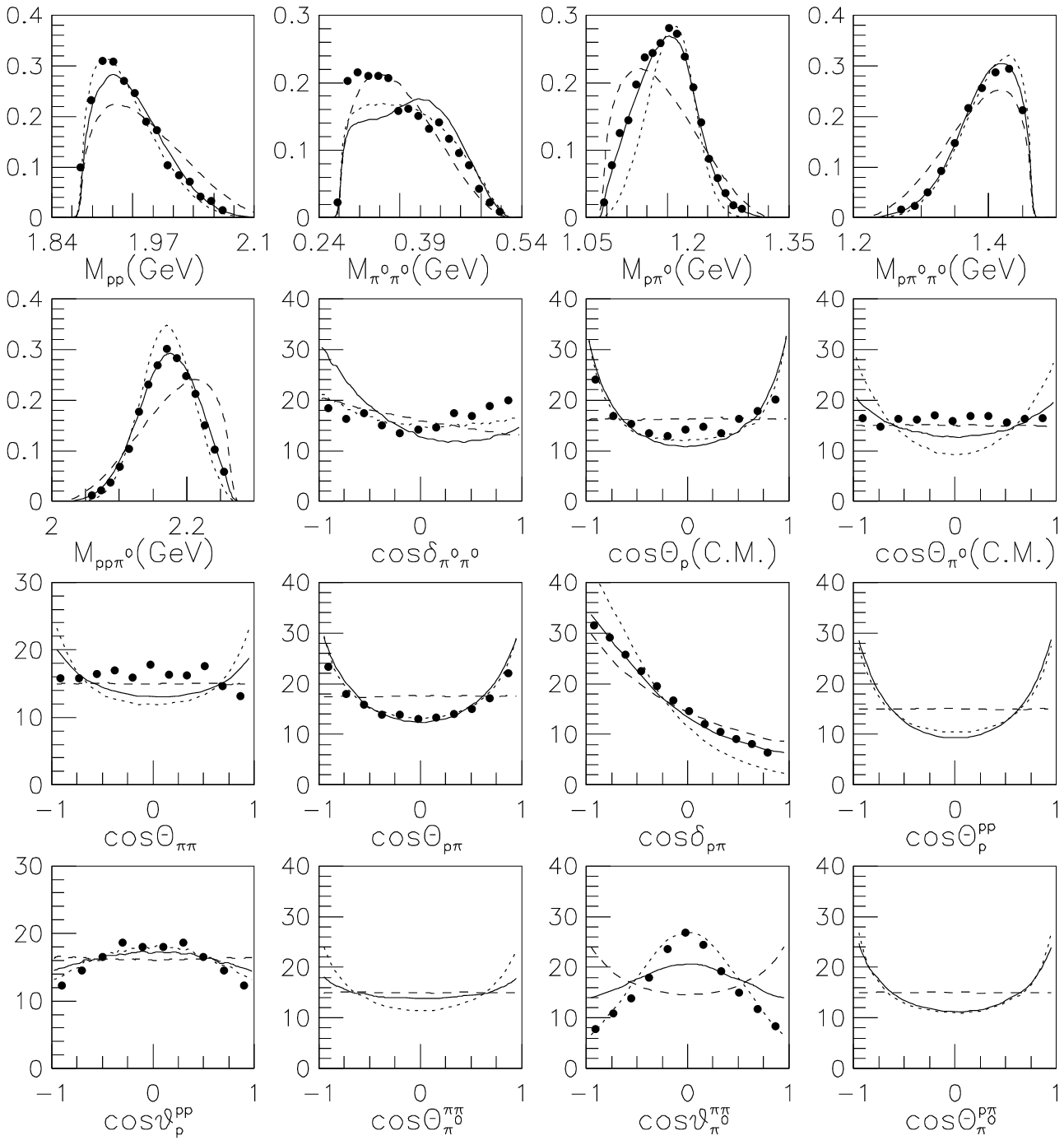}}
    \caption{Differential cross sections of $pp \to pp\pi^0\pi^0$
    at beam energies 1200 MeV. The meaning of curves is the same as
    Fig.~\protect\ref{2pi01000}. The data are from
    Ref.~\protect\cite{skorodko}.} \label{2pi01200}
  \end{center}
\end{figure}
\begin{figure}[htbp]
  \begin{center}
{\includegraphics*[scale=1.0]{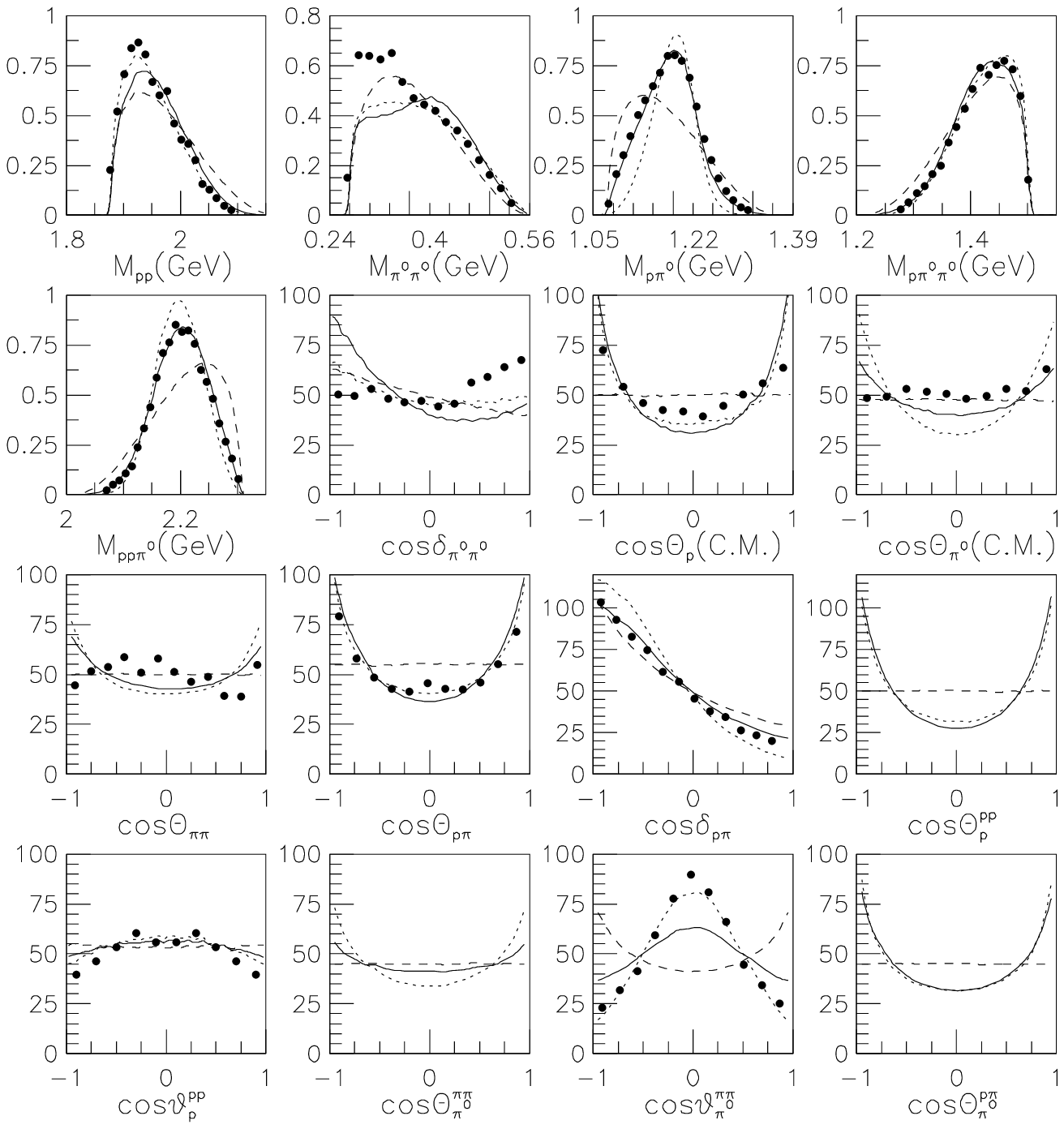}}
    \caption{Differential cross sections of $pp \to pp\pi^0\pi^0$
    at beam energies 1300 MeV. The meaning of curves is the same
    as Fig.~\protect\ref{2pi01000}. The data are from
    Ref.~\protect\cite{skorodko}.} \label{2pi01300}
  \end{center}
\end{figure}
\begin{figure}[htbp]
  \begin{center}
{\includegraphics*[scale=0.7]{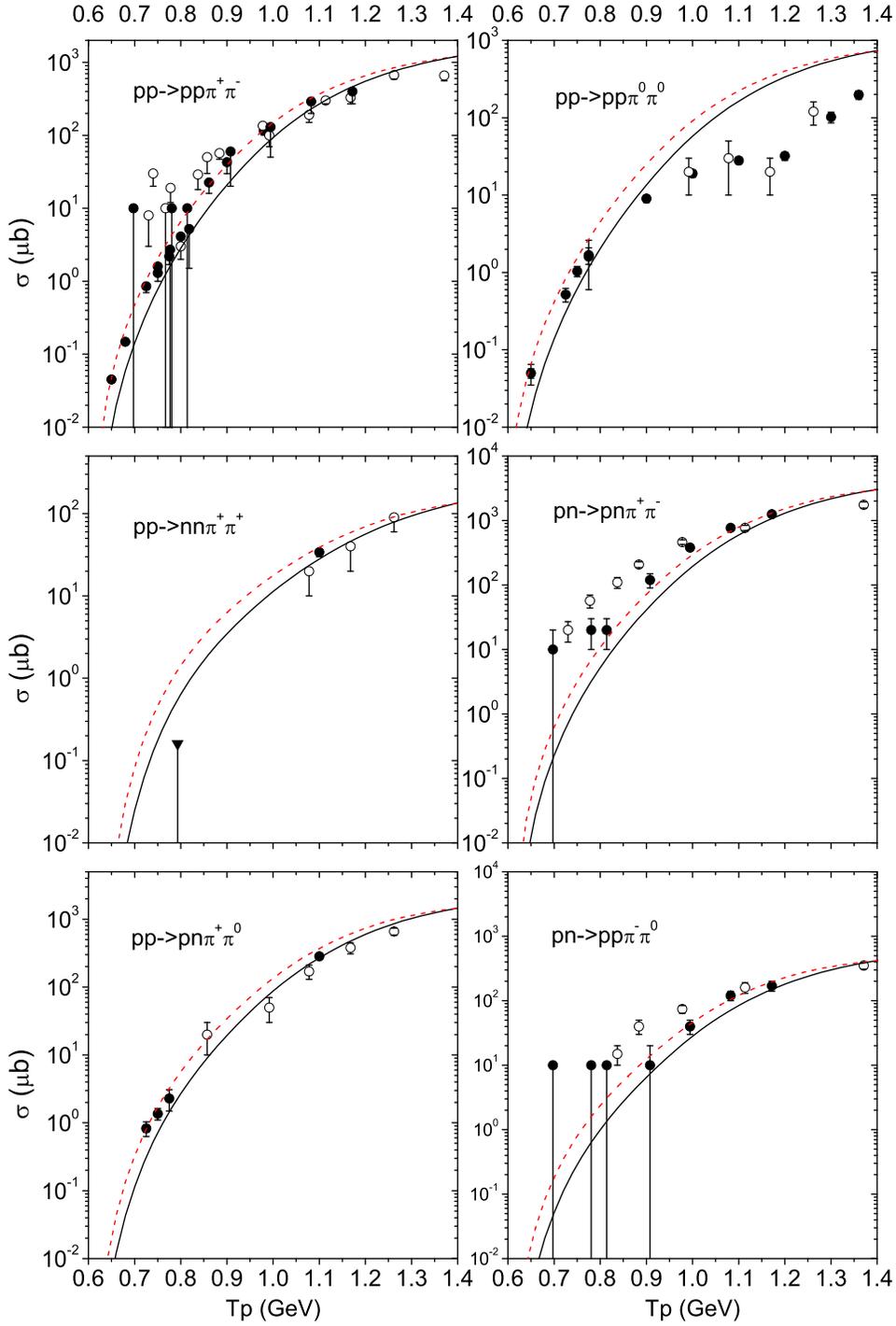}}
    \caption{(color online) Total cross sections of $NN \to NN\pi\pi$.
    The black solid and red dashed curves correspond to the full
    contributions without and with final state interactions,
    respectively. The data are the same as the Fig.~\protect\ref{tcs}}\label{tcsfsi}
  \end{center}
\end{figure}
\begin{figure}[htbp]
  \begin{center}
{\includegraphics*[scale=0.7]{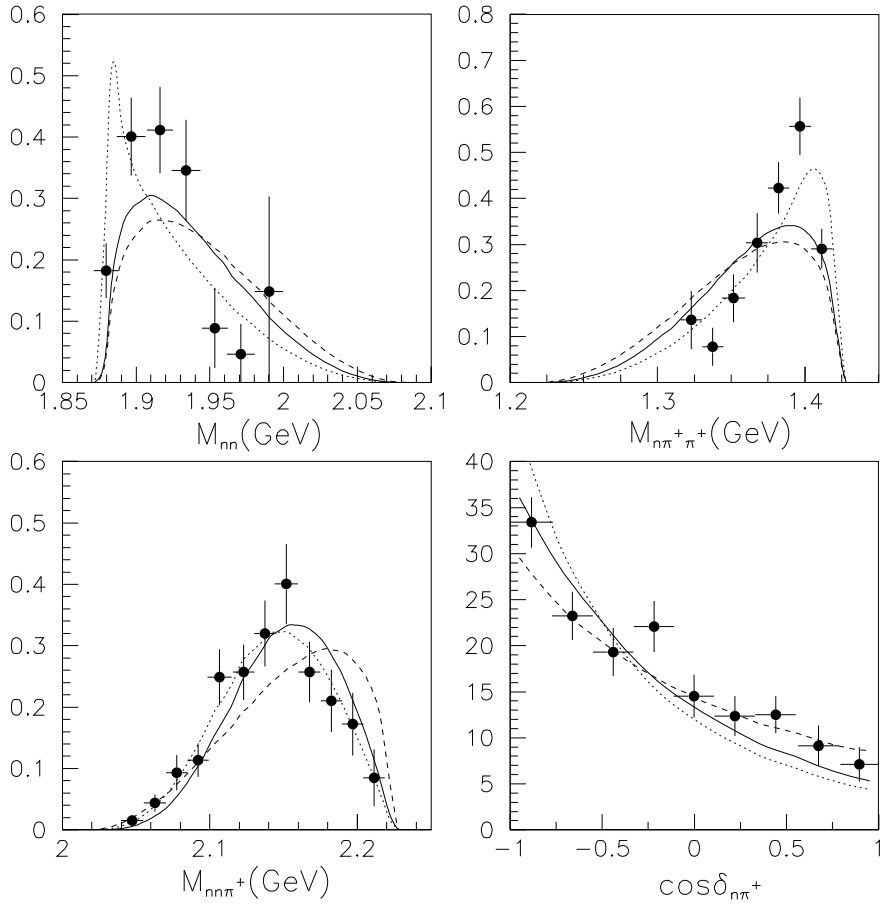}}
    \caption{Differential cross sections of $pp \to nn\pi^+\pi^+$
    at beam energies 1100 MeV. The dashed, dotted and solid curves
    correspond to the distributions of the phase space, the full
    model with and without FSI, respectively. The data are from
    Ref.~\protect\cite{dataIJMPA}.} \label{nnpipif}
  \end{center}
\end{figure}

\end{document}